\documentclass{aa}
\usepackage{epsf,amsfonts,amssymb,graphicx,fancyheadings,caption} 
\usepackage{multirow}
\begin{document} 


\title{Bulge Microlensing Optical Depth from EROS~2
  observations\thanks{Based on observations made with the MARLY
  telescope at the European Southern Observatory, La Silla, Chile.}}
\author{
C.~Afonso\inst{1,5,11},
J.N.~Albert\inst{2},
C.~Alard\inst{3},
J.~Andersen\inst{7},
R.~Ansari\inst{2},
\'E.~Aubourg\inst{1},
P.~Bareyre\inst{1,5},
F.~Bauer\inst{1},
J.P.~Beaulieu\inst{6},
G.~Blanc\inst{1}, 
A.~Bouquet\inst{5},
S.~Char\inst{8}\thanks{deceased},
X.~Charlot\inst{1},
F.~Couchot\inst{2},
C.~Coutures\inst{1},
F.~Derue\inst{1,2}\thanks{Presently at Centre de Physique des
  Particules de Marseille, IN2P3-CNRS, 163 Avenue de Luminy, case 907,
  13288 Marseille Cedex 09, France},
R.~Ferlet\inst{6},
P.Fouqu\'e\inst{10,12}
J.F.~Glicenstein\inst{1},
B.~Goldman\inst{1,5,11},
A.~Gould\inst{5,9},
D.~Graff\inst{9},
M.~Gros\inst{1},
J.~Haissinski\inst{2},
C.~Hamadache\inst{1},
J.C.~Hamilton\inst{5}\thanks{Presently at ISN, IN2P3-CNRS-Universit\'e
  Joseph-Fourier, 53 Avenue des Martyrs, 38026 Grenoble Cedex, France},
D.~Hardin\inst{1}\thanks{Presently at LPNHE, IN2P3-CNRS-Universit\'e
  Paris VI et VII, 4 Place Jussieu, F-75252 Paris Cedex, France},
J.~de Kat\inst{1},
A.~Kim\inst{5}\thanks{Lawrence Berkeley National Laboratory, Berkeley,
  CA 94720, U.S.A.}, 
T.~Lasserre\inst{1},
L.LeGuillou\inst{1},
\'E.~Lesquoy\inst{1,6},
C.~Loup\inst{6},
C.~Magneville\inst{1},
B.~Mansoux\inst{2},
J.B.~Marquette\inst{6},
\'E.~Maurice\inst{4},
A.~Maury\inst{12},
A.~Milsztajn \inst{1},
M.~Moniez\inst{2},
N.~Palanque-Delabrouille\inst{1},
O.~Perdereau\inst{2},
L.~Pr\'evot\inst{4},
N.~Regnault\inst{2},
J.~Rich\inst{1},
M.~Spiro\inst{1},
P.~Tisserand\inst{1},
A.~Vidal-Madjar\inst{6},
L.~Vigroux\inst{1},
S.~Zylberajch\inst{1}
\\   \indent   \indent
The EROS collaboration
}    
\institute{
CEA, DSM, DAPNIA,
Centre d'\'Etudes de Saclay, 91191 Gif-sur-Yvette Cedex, France
\and
Laboratoire de l'Acc\'{e}l\'{e}rateur Lin\'{e}aire,
IN2P3 CNRS, Universit\'e de Paris-Sud, 91405 Orsay Cedex, France
\and
DASGAL, INSU-CNRS, 77 avenue de l'Observatoire, 75014 Paris, France
\and
Observatoire de Marseille,
2 pl. Le Verrier, 13248 Marseille Cedex 04, France
\and
Coll\`ege de France, Physique Corpusculaire et Cosmologie, IN2P3 CNRS,
11 pl. M. Berthelot, 75231 Paris Cedex, France
\and
Institut d'Astrophysique de Paris, INSU CNRS,
98~bis Boulevard Arago, 75014 Paris, France
\and
Astronomical Observatory, Copenhagen University, Juliane Maries Vej 30,
2100 Copenhagen, Denmark
\and
Universidad de la Serena, Facultad de Ciencias, Departamento de Fisica,
Casilla 554, La Serena, Chile
\and
Department of Astronomy, Ohio State University, Columbus,
OH 43210, U.S.A.
\and 
Observatoire de Paris, LESIA, 92195 Meudon Cedex, France
\and
Department of Astronomy, New Mexico State University, Las Cruces, NM
88003-8001, U.S.A.
\and
European Southern Observatory (ESO), Casilla 19001, Santiago 19, Chile
}
\offprints{afonso@hep.saclay.cea.fr}         
\date{Received;accepted}

\authorrunning{C. Afonso et al.}
\titlerunning{Bulge Microlensing Optical Depth from EROS~2 observations}

\def\lsim{{\lesssim}}
\def\au{{\rm AU}}
\def\etal{{et al.}}
\def\eros{{\sc eros}}
\def\macho{{\sc macho}}
\def\lmc{{\sc lmc}}
\def\smc{{\sc smc}}
\def\ie{{\em i.e.}}
\def\tempest%
{\begin{array}{ccc}
1 & 1 & 1 \\
1 & 1 & 1 \\
4 & 3 & 8
\end{array}}
\def\gsim{{{}_>\atop{}^{{}^\sim}}}
\def\lsim{{{}_<\atop{}^{{}^\sim}}}
\def\kms{{\rm km}\,{\rm s}^{-1}}
\def\kpc{{\rm kpc}}
\def\e{{\rm E}}
\def\rel{{\rm rel}}
\def\btheta{{\vec\theta}}
\def\bmu{{\vec\mu}}
\def\bpi{{\vec\pi}}
\def\Teff{{T_{\rm eff}}} 
\def\msun{\rm M_\odot} 

\abstract{ 
        We present a measurement of the microlensing optical depth toward the
Galactic bulge based on the analysis of 15 contiguous
$1\,\rm deg^2$ fields centered on $(l=2.\hskip-2pt^\circ 5,
b=-4.\hskip-2pt^\circ 0)$ and containing 
$N_*=1.42\times 10^{6}$ clump-giant stars (belonging to the extended
clump area) monitored during almost three
bulge seasons 
by EROS (Exp\'erience de Recherche d'Objets Sombres). We find
$\tau_{bulge}=0.94\pm 0.29\times 10^{-6}$ averaged over all fields,
based on 16 microlensing events with clump giants as sources. This
value is substantially below several other determinations by the MACHO
and OGLE groups and is more in agreement with what is expected from
axisymmetric and non-axisymmetric bulge models.  

\keywords{Galaxy:bar - Galaxy:stellar contents - Galaxy:structure -
  Cosmology:gravitational lensing}  
} 

\maketitle           

\section{Introduction} 
\label{section:introduction}

        When microlensing surveys toward the Galactic bulge were first
proposed by Paczy\'nski (1991) and Griest (1991), it was 
expected that the optical depth to microlensing in the Baade Window
($l=1^\circ,b=-3.\hskip-2pt^\circ9$) due to ordinary disc lenses would
be $\tau \sim 4\times 10^{-7}$. In the presence of brown dwarfs in the disc
with a total mass density equal to that of ordinary stars, the 
microlensing optical depth toward the bulge would rise to $\tau \sim
8\times 10^{-7}$. The initial detections by OGLE reporting six
microlensing events (\cite{UDA94a}) seemed to indicate that the
optical depth was higher
than the predicted values, although no  estimate was published. Kiraga
\& Paczy\'nski (1994) 
then realized that the contribution of lenses in the bulge had also to
be considered and that the density of disc lenses had to be reevaluated. 
They concluded that the bulge itself  
would most likely dominate the event rate.
Nevertheless, when OGLE obtained an optical depth in the Baade Window
of  $\tau=(3.3\pm1.2)\times 10^{-6}$ with a sample of 9 microlensing events
(\cite{UDA94b}) and MACHO made the first  
formal estimate $\tau_{bulge}=3.9^{+1.8}_{-1.2}\times 10^{-6}$ (at
$l=2.\hskip-2pt^\circ55, b=-3.\hskip-2pt^\circ64$) based on 13 events
with clump-giant 
sources (\cite{ALC97}), the community found it quite surprising. In
the same paper MACHO also derived 
$\tau = 2.4\pm 0.5\times 10^{-6}$, based on 41 events, including not
only clump giants, but all sources. They argued, however, that 
the determination of the optical depth for fainter
source stars is less straightforward than for bright ones due to 
blending problems in crowded fields, where a source star can be a blend of
two or more stars. Hence the entire luminosity function has to be
modeled to account for both resolved and unresolved sources.   

Gould (1994) and Kuijken (1997) showed that the expected maximum
optical depth generated by 
axisymmetric mass distributions of the Galaxy was surpassed by the
observations. Indeed, attention was immediately focused on the
possibility that the high microlensing rate represented yet another
detection of a (non-axisymmetric) bar in the central regions of the
Galaxy. At this time, a ``bar consensus'' was developing based on gas
kinematics (\cite{BIN91}), infrared light measurements
(\cite{DWEK95}), and star counts (\cite{NIK97}). However, even
barred bulge models, with various values for the bar mass and the
orientation to our line of sight, predict optical depths
systematically lower than the    
observed values: Han \& Gould (1995b) found $1.5\times
10^{-6}< \tau_{bulge} <2\times 10^{-6}$ at the Baade Window for
bulge-giant sources; Zhao et al. (1996) determined 
$\tau_{bulge}=2.2\times 10^{-6}$ for clump-giant sources at 
the MACHO field positions ($l=2.\hskip-2pt^\circ.55,
b=-3.\hskip-2pt^\circ64$); 
Zhao \& Mao (1996) showed 
that several boxy and ellipsoidal-type bar models constrained by the
COBE maps produce optical depths at the Baade Window $2\sigma$ lower
than MACHO and OGLE measured values, even with a massive bar
$M_{bar}=2.8\times10^{10}\msun$ and a small orientation angle
$\theta<20^\circ$ to the line of sight. Moreover, Binney, Bissantz \&
Gerhard (2000) recently showed that $\tau\sim 4\times 10^{-6}$
cannot be produced by any plausible non-axisymmetric model of the Galaxy.

        Up to the present there have been several other estimates of $\tau$.
Alcock et al. (2000) analyzed a subset of
three years of MACHO data using difference imaging.  This method
increases the number of detected  
microlensing events. The mean optical depth (to the heterogeneous
collection of bulge and disc sources in the MACHO fields) based on the 99
events found by this technique is estimated to be 
$\tau=2.43^{+0.39}_{-0.38}\times 10^{-6}$ at
($l=2.\hskip-2pt^\circ 68, b=-3.\hskip-2pt^\circ 35$). MACHO corrected
this value to the true optical depth to bulge sources by assuming that
25\% of the sources lay in the foreground and therefore did not
contribute significantly to the observed microlensing events. They
found the optical depth to be $\tau_{bulge}=3.23^{+0.52}_{-0.50}\times
10^{-6}$. Another optical depth value is given 
by Popowski et al. (2000). Their analysis of 5 years of
MACHO data revealed 52 microlensing events with clump-giant
sources. The corresponding optical depth is
$\tau_{bulge}=(2.0\pm0.4)\times 10^{-6}$ averaged 
over 77 fields centered at ($l=3.\hskip-2pt^\circ 9,
b=-3.\hskip-2pt^\circ 8$).      
A large fraction, perhaps a majority, of events detected toward
the bulge have been found by OGLE~II (\cite{UDA00}, \cite{WOZ01})
but so far these have not been used to estimate $\tau$, as the
OGLE~II experimental detection efficiencies, necessary for the
determination of microlensing optical depths, have not been made
available yet.

        In this paper we present the first estimate of the EROS~2 
optical depth to microlensing toward the Galactic Center. The EROS~2
bulge survey, begun in July 1996, was specifically designed to  
find events with bright sources including the extended clump area
(see Fig. \ref{fig:cmd})and other giants 
because, as discussed above, these can be interpreted unambiguously
(\cite{GOU95b}).

\section{Data}
\label{section:data}

        The data were acquired at the EROS~2 team 1 m MARLY telescope
at La Silla, Chile. The imaging was done simultaneously by two cameras,
using a dichroic beam-splitter. Each camera is composed of a mosaic of 
eight 2K$\times$2K LORAL CCDs, with a pixel size of $0.\hskip-2pt''6$ and a
corresponding field of $0.\hskip-2pt^\circ 7(\alpha)\times
1.\hskip-2pt^\circ 4(\delta)$.  
One camera observes in the so-called EROS-blue filter and the other
in the so-called EROS-red filter, these filters having been
specifically designed to cover a broad passband to 
collect as many photons as possible. Thus, the EROS filters are
non-standard: EROS-red (620-920 nm) is roughly equivalent to Cousins
$I$, but larger, while EROS-blue (420-720 nm) is a
band that overlaps Johnson/Cousins $V$ and $R$. Details about the
instrument can be found in Bauer et al. (1997). For information about the
acquisition pipeline see Palanque-Delabrouille (1997).    

\begin{figure}[h]
 \centering
 \includegraphics[width=7.8cm]{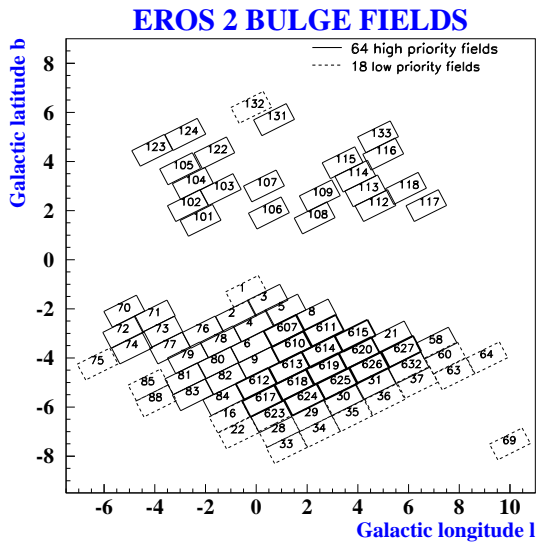}
 \caption{Galactic plane map of the EROS 2 bulge fields. A total of 82
   fields are monitored. The high priority fields are supposed to be
   observed at least every other night (solid lines). The
   lower priority fields (dashed lines) are only monitored if some
   observation time is still available after the high prioriy
   sequence. The analysis and the results reported in this paper
   concern 15 deg$^2$ (bold lines).}
  \label{fig:eros2_bulge_fields}
\end{figure}

        Although the total sky area covered by the EROS bulge survey is
$82\,\rm deg^2$, the observations reported here concern only 15 of these
fields, monitored between mid-July 1996 and 31 May
1999. Fig.\ref{fig:eros2_bulge_fields} shows the location of the 82 fields in
the galactic plane $(l,b)$. We also indicate the fields classified as
high-priority (solid line), with the largest number of red clump
giants, which we attempt to observe every other night. The 
lower priority fields (dashed line) are monitored only if there is
still enough time left after the high-priority sequence, taking into
account the compromise between the bulge  
survey and other EROS targets: Spiral Arms (\cite{DER01}), LMC
(\cite{LASS00}), SMC (\cite{AFO99}), proper motion survey
(\cite{GOLD02}), supernova search (\cite{HAR00}). The 15
fields whose analysis is presented here are marked in bold. The
corresponding data set contains $2.3\times 10^6$ light curves, of which
$1.4\times 10^6$ are bulge clump giants of the
  extended clump area (see Fig. \ref{fig:cmd}). As we mentioned above,
our bulge program was specifically designed to select events with
bright stars as sources to avoid blending problems. Since one of the
red CCDs was not functioning well during a large fraction of the observation
period, it was not included in our analysis, since we require two colours.

\section{Event Selection}
\label{section:event_selection}

The image photometry was performed with software specifically designed for
crowded fields, PEIDA (Photom\'etrie et \'Etude d'Images Destin\'ees
\`a l'Astrophysique) (\cite{ANS96}). After the production of the
light curves and removal of defective data points due to images with a 
specific problem (bad atmospheric conditions, temporary instrumental 
deficiencies), several cuts were applied to the data set. The
selection criteria explained in detail below are based on the
characteristics of microlensing events light curves, which follow
the Paczy\'nski (1986) function  
\begin{eqnarray} 
F(t) & = & F_{\rm s} A[u(t)] \\
{\rm where\;\;\;} A(u) & = & \frac{u^{2}+2}{u\sqrt{u^{2}+4}} \\
{\rm and\;\;\;\;}u^{2}(t) & = & u_{0}^{2} + \frac{(t-t_0)^2}{t_\e^2} \; .
\end{eqnarray} 
These equations contain 5 parameters (which are obtained for each star
by fitting the Paczy\'nski profile to the corresponding light curve):
two baseline fluxes ($F_{\rm s}$) of the source star in the red and blue
EROS filters, $F_{\rm R_{EROS}}$ and $F_{\rm B_{EROS}}$, the date of maximum
amplification $t_0$, the
impact parameter $u_0=u(t_0)$ (i.e. the minimum lens-source separation
projected in the lens plane, normalized by the Einstein ring radius
$R_\e$) and finally, the microlensing event duration, i.e. 
the Einstein ring radius crossing time $t_\e=R_\e/v_{\rm t}$. The time
scale $t_\e$ 
contains a 3-fold degeneracy between the transverse velocity $v_{\rm t}$ of
the lens, its mass $M$, and its distance from the observer $D_{\rm OL}$.
The $t_\e$ dependence on the mass and the distance comes through
$R_\e^2=4GMD_{\rm LS}D_{\rm OL}/c^2D_{\rm OS}$, where $D_{\rm OS}$ is
the distance between the observer and the source and $D_{\rm LS}$ the
distance between the lens and the source. 

    The main characteristics of an amplified light curve of a source
star during gravitational microlensing are a symmetric shape
in time, with an increasing light intensity as the foreground lens
approaches the line of sight to the background source star, and then
decreases as the lens moves away (assuming a constant transverse
velocity $v_t$ of the lens). When blending is
neglected, the amplification $A[u(t)]$ is the same in the two
observing bands (the imaging being done simultaneously in both bands)
and therefore achromatic, since microlensing is a purely 
geometrical phenomenon and should thus not depend on the observing
wavelength. When the reconstructed star is a blend of two or more
stars, the observed baseline flux $F_{\rm s}=F_1+F_2$ is the sum of
the flux $F_1$ of the main component of star and the flux $F_2$ due to
unresolved background stars. Assuming that the main component is amplified by a factor
$A(t)$, the observed flux during the amplification is given by
$F(t)=F_1A(t)+F_2=F_{\rm s}A_{\rm obs}(t)=ACF_{\rm s}A(t)+(1-C)F_{\rm
  s}$ where $C=(A_{\rm obs}(t)-1)/(A(t)-1)$ is the blending
coefficient and $A_{\rm obs}(t)$ is the observed amplification.  

     Finally, another main characteristic is that the  
amplification peak should be unique for a given source star, as the
probablitity of a star to be lensed is extremely low, of the order of
one per $10^6$ stars. If two or more variations occur, the source
star is more likely to be variable. Thus, several of our cuts
concern the rejection of variable stars. 

   Hereafter we describe the selection criteria, which
are similar to those used in the other EROS microlensing programs.

\begin{enumerate}
\item {\bf Rejecting stable stars}\\ 
\begin{enumerate}
\item The main fluctuations in both the red and blue light curves
  should be positive\footnote{A positive (negative) fluctuation is
  defined as a set of consecutive points that deviate by at least
  1$\sigma$ above (below) the baseline flux.}, and should overlap
  in time by at least 10\%. 
\item To select light curves with a significant main fluctuation
    we use the discriminant $LP$\footnote{The statistical significance of a fluctuation
    containing $N$ points is given by $LP = -
    \sum_{i=1}^{i=N}\log\left(\frac{1}{2}\:{\rm
        Erfc}\left(\frac{x_{i}}{\sqrt{2}}\right)\right)$, with each 
     point $i$ deviating at time $t_i$ by $x_i$ (in units of
     $\sigma_i$, the error for the i-th flux measurement).}, to
     which no true statistical meaning is assigned. It is rather used
     in an empirical and relative manner, in the sense that the light
     curves with higher $LP$ values 
     than other light curves have a more significant variation. Thus
   we require that the main fluctuation in both colours is 
   significant: $LP(\rm main\;fluct.)>40$.\\
\end{enumerate}

\vspace{0.5cm}

\item {\bf Eliminating variable stars}\\ 
\begin{enumerate}  
\item To reject the scattered light curves of short period
  variable stars, which vary on time scales shorter than the average time
  sampling of our fields, the
  following requirement is made: the distribution of the
  difference, in units of $\sigma_{\rm i}$ (the error for the i-th flux
  measurement) , between each flux $F(t_{\rm i})$ and the 
  linear interpolation of the two adjacent neighbors $F(t_{\rm i-1})$ and
  $F(t_{\rm i+1})$, should have an RMS lower than 2.5. 

\begin{displaymath}
   \sigma_{int} < 2.5 \;, 
\end{displaymath}
 
\item Longer period variable stars display variations in both 
  red and blue bands. They are likely to show such correlated variations
  outside the principal fluctuation. Such correlations are searched for 
  using the Fisher variable (FV) which is a function of the
  correlation coefficient $\rho$ between the red and blue fluxes. This
  variable allows one to distinguish, with a  
  better resolution, between correlation values very close to each
  other and thus to tune more precisely the cut. We require  

\begin{displaymath}
   FV(\rho) = 0.5\times\sqrt{N-3}\times
 \ln{\left(\frac{1+\rho}{1-\rho}\right)} < 13\;, 
\end{displaymath}
 
  where $N$ is the number of pairs of simultaneous measurements in the red
  and blue bands, belonging to the unamplified part of the
  light curve. The exclusion of the principal fluctuation (plus a security
  time margin) 
  guarantees the survival of the microlensing candidates, which
  as expected exhibit a strong correlation within the amplification peak. 
\item The following rejection criterion is similar to 2(a) and
  2(b), eliminating the variable stars that passed these cuts. We
  keep only the light curves that have a stable baseline outside
  the principal fluctuation in both bands

\begin{displaymath}
  \chi^2(baseline)=\frac{\chi^2_{\rm ml}(baseline)}{d.o.f.(baseline)}<5\;,   
\end{displaymath}   

  where $\chi^2_{ml}(baseline)$ and $d.o.f.(baseline)$ are
  respectively the chi square of the microlensing fit (carried out
  separately in each band) 
  and the number of degrees of freedom of the fit, both values
  concerning the unamplified part of the light curve.\\ 
\end{enumerate}
\item {\bf Selecting high S/N events}\\
\begin{enumerate}
\item To select events with a high signal-to-noise ratio (S/N) a
  cut is applied to a semi-empirical estimator, whose value will
  increase as a microlensing fit (ml) improves over a constant-flux
  fit (cst)

\begin{displaymath}
 \Delta\chi^2 = \frac{\chi^2_{\rm cst} -
  \chi^2_{\rm ml}}{\chi^2_{\rm ml}/d.o.f.}\frac{1}{\sqrt{2 d.o.f.(peak)}}\;, 
\end{displaymath}

 where $d.o.f.$ is the number of degrees of freedom of the fit
 over the entire light curve and $d.o.f.(peak)$ refers to the
 number of degrees of freedom of the fit within the amplification
 peak. For the fits we use simultaneously the data points of both red
 and blue light curves. We require $\Delta\chi^2>70$. 
\item Candidates with low fitted maximum amplifications
  $A(u_0)$, may be due to statistical fluctuations or systematic
  photometry biases, or may be
  impossible to distinguish from these when the photometric 
  precision of the stars (which for clump giants is of the order of
  2-3\%) does not allow it. To remove these candidates from the
  remaining set, we demand for each star that its maximum
  amplification be  
  greater than 5 times the photometric precision of the star
  (calculated from the unamplified part of the light curve). For a
  2-3\% photometric precision, this cut allows the detection of 
  maximum amplifications as low as 10\%.\\    
\end{enumerate}
\item {\bf Date of maximum amplification and time span allowing to validate a candidate}\\
\begin{enumerate}
\item Although the above criteria select candidates that
  {\sl a priori} are microlensing events, some exhibit their
  date of maximum amplification $t_0$ just before or after the
  observation period. The confirmation of a candidate for which we have
  only the decreasing or increasing part of the amplification peak on
  the light curve is extremely difficult, if not impossible. Thus, we
  require that the fitted date of maximum amplification $t_0$ is
  within the observation period

\begin{displaymath}
  T_{\rm first}-\frac{t_{\e}}{3}<t_{0}<T_{\rm last}+\frac{t_{\e}}{3}\;,
\end{displaymath}

  where $t_\e$ is the event time scale, $T_{first}$ is the first day of
  the observations and $T_{last}$ the last one. A margin is allowed
  ($t_{E}/3$) due to the uncertainty of the fitted $t_0$ value. 
\item As for the previous cut, it is also difficult to
  confirm candidates with time scales $t_\e$ too long compared to the
  observation period ($T_{\rm obs}\sim 3$ years), even if the date of
  maximum amplification is contained in the light curve. We demand
  that the observation period be at least 3 times greater than the
  Einstein ring radius crossing time $t_\e$, so that the starting or ending 
  points of the amplification are visible on the light curves. The
  candidates removed by this cut from the final set (with $t_\e>400$
  days), are kept on a list for regular follow-up and checking, as they
  could be due to black holes or neutron stars.   
\end{enumerate}
\end{enumerate}

    To be chosen as a candidate, the light curve must satisfy each
one of the above listed criteria. These are tuned by applying the same
selection criteria to the data and to a set of simulated microlensing
events (generated on top of the real light curves, see
\S\ref{section:efficiencies}). One tries to eliminate a 
maximum of false candidates, while keeping the greatest possible
number of simulated events. In order to detect also non-standard
microlensing events (source size effect, caustic crossing), the
selection criteria have been tuned sufficiently loosely.

\section{EROS~2 bulge microlensing candidates} 
\label{section:candidates}

\begin{figure}[h]
 \centering
 \includegraphics[width=7.8cm]{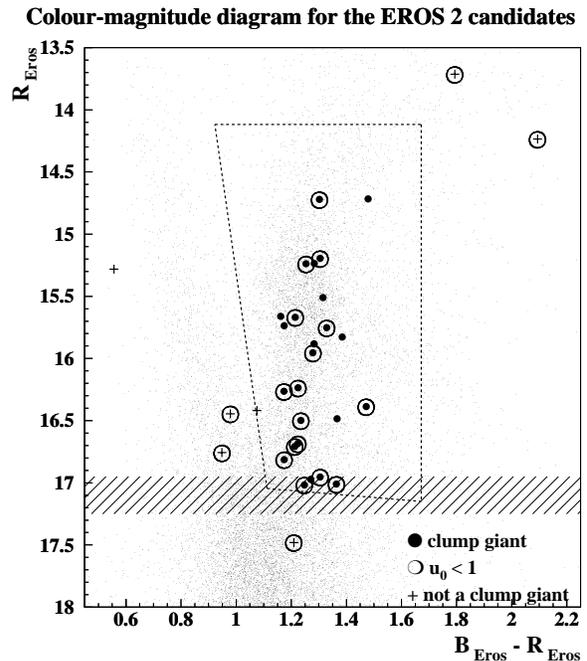}
 \caption{Colour-magnitude diagram (CMD) for the source stars of the EROS~2
   microlensing candidates superimposed (after alignment) on the CMD
   of the stars of the EROS sub-field centered on the Baade
   Window. Clump-giant sources (of the extended clump area) of the
   microlensing candidates are indicated with solid  
   circles. Sources other than clump giants are marked with
   crosses. The empty circles surrounding the markers refer to
   microlensing candidates with a maximum amplification
   $A_0>1.34$. The magnitude cut in the CMD for the selection of
   bright reference source stars varies from field to field. This
   variation is indicated by the hatched area. Finally, the dashed lines
   surround the extended clump area containing the stars used for the
   optical depth calculation.}  
  \label{fig:cmd}
\end{figure}

The selection criteria presented in the previous section yield a
total of 33 microlensing candidates, of which 25 have clump-giant 
sources (belonging to the extended clump area). Fig.~\ref{fig:cmd} shows
the location of the  
source stars for the 33 microlensing candidates in an instrumental
colour-magnitude diagram (CMD). This diagram was obtained by splitting
up the EROS $0.7\times1.4\;{\rm deg}^2$  field into 32
$0.17\times0.17\;{\rm deg}^2$ sub-fields, finding 
the center of the clump of each of the sub-fields and then aligning
them independently to an
arbitrary common position on an instrumental CMD, which was chosen to
be the EROS field centered on the Baade Window. To define $R_{\rm EROS}$
and $B_{\rm EROS}$ magnitudes, stars in the OGLE Baade Window catalog
(with field coordinates
$\alpha(J2000)=18^h03^m37,\delta(J2000)=-30^\circ05'00''$)  
(\cite{PAC99}) were matched with EROS stars  

\begin{eqnarray}
R_{\rm Eros} & = & 26.95 - 2.5\times\log(F_{\rm R_{\rm EROS}}) \\
B_{\rm Eros} & = & 27.86 - 2.5\times\log(F_{\rm B_{\rm EROS}}) \; .
\end{eqnarray}
where $F_{\rm R_{\rm EROS}}$ and $F_{\rm B_{\rm EROS}}$ are the red and blue fluxes (in
ADU/120 s) of the center of the clump in the EROS sub-field corresponding to
the Baade Window. 
The source stars of the microlensing candidates believed to be clump
giants of the extended clump area are marked with solid circles, and
sources other than clump giants are 
depicted with crosses. The markers surrounded 
by open circles refer to microlensing candidates with a maximum
amplification $A_0>1.34$, i.e. an impact parameter $u_0<1$. The
hatched area indicates the variation from field to field of the CMD
adopted apparent magnitude cuts. Indeed, as we already mentioned in
\S\ref{section:introduction}, the 
purpose of the EROS bulge program was to find events with bright
sources so as to avoid uncertainties due to blending. The selection of
these sources was  
made by determining the center of the clump in the CMD of each sub-field
with a special search algorithm, and rejecting all the stars below the
lower limit of the clump minus  0.5 magnitude. The lower limit of the
clump is defined as being 1.5$\sigma$ away from the mean of a Gaussian
fitted along the magnitude axis of the CMD. Finally, the dashed
lines delimit the extended clump area.  

\begin{figure}[h]
\begin{minipage}[c]{.49\textwidth}
 \includegraphics[width=7.5cm]{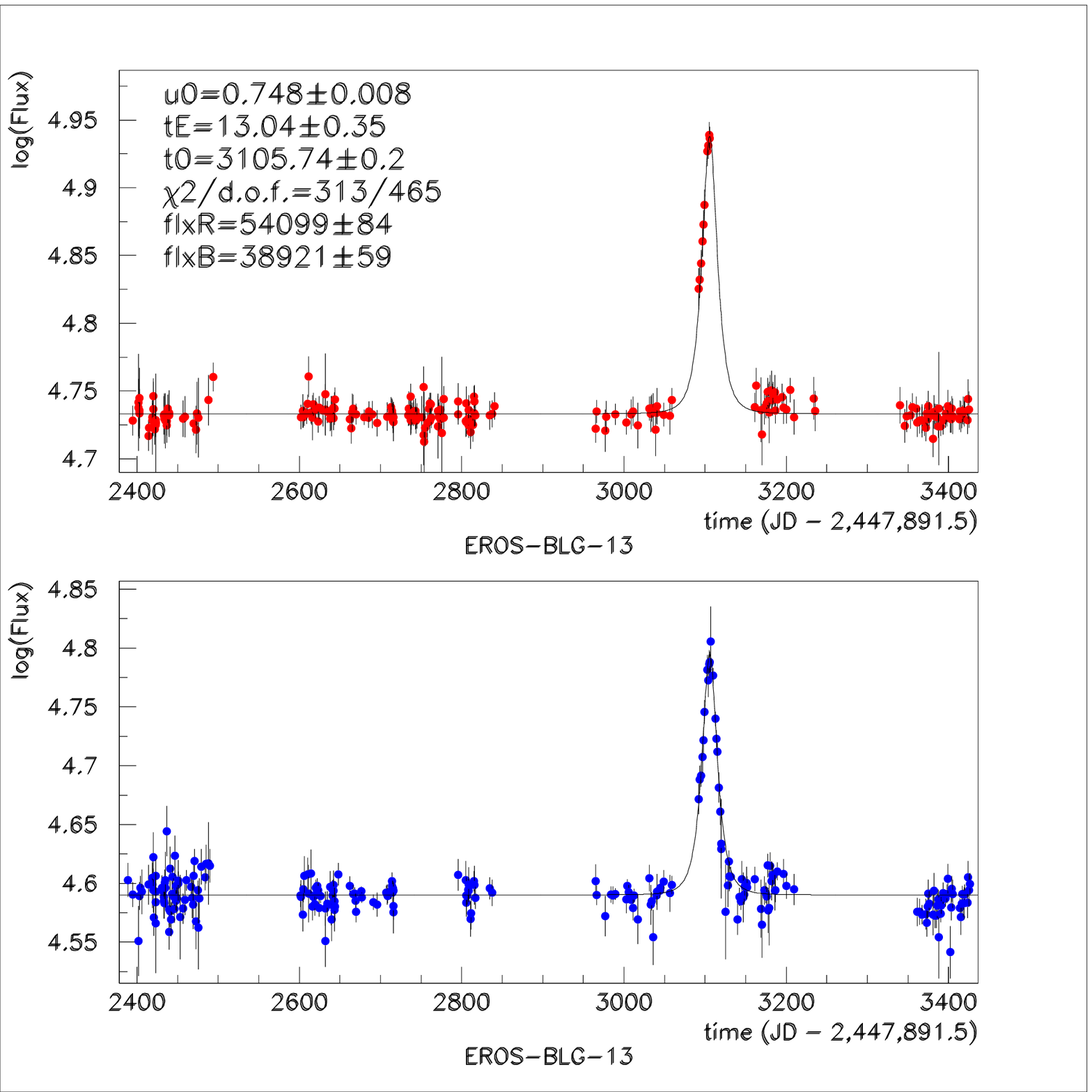}
\end{minipage}
\begin{minipage}[c]{.49\textwidth}
 \includegraphics[width=7.5cm]{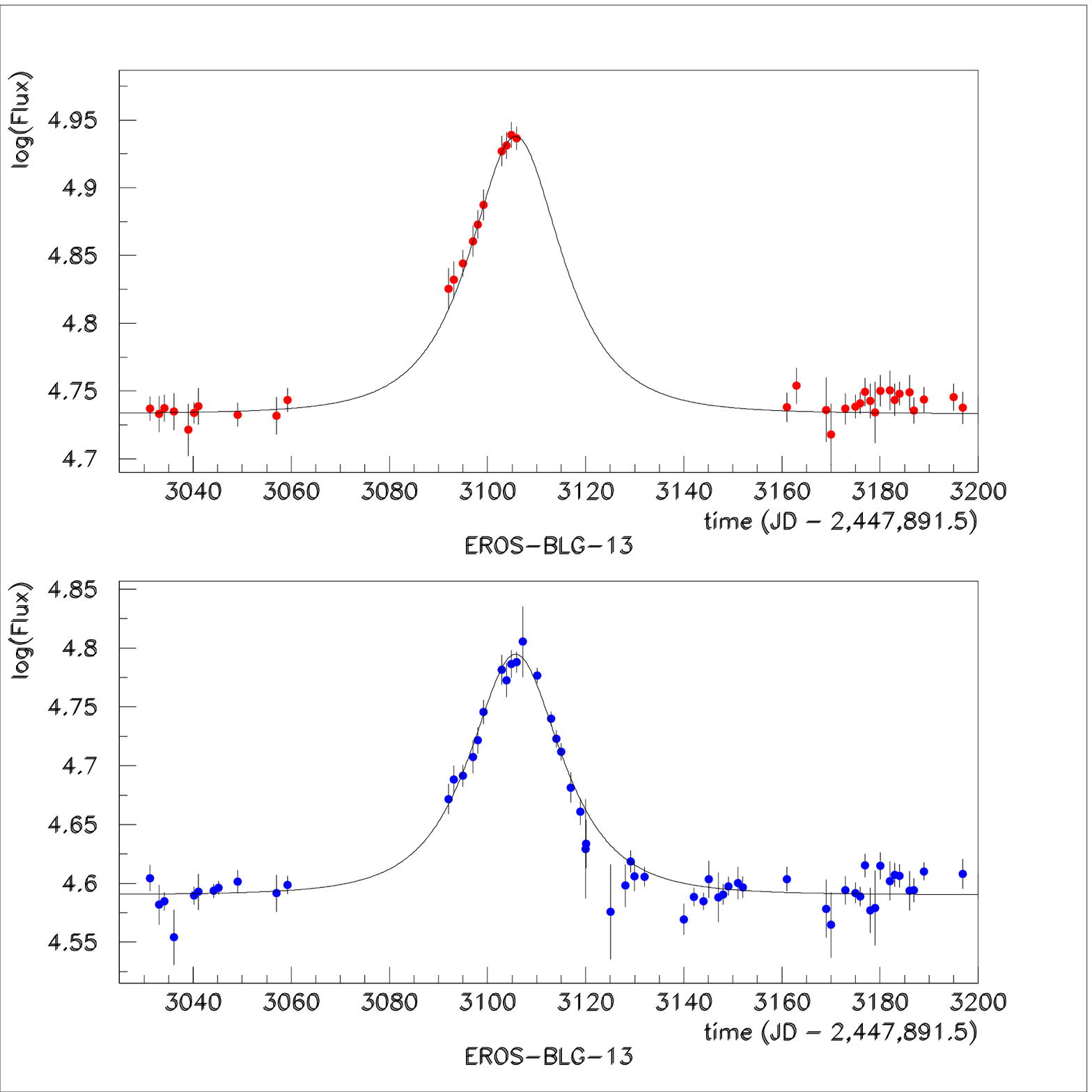}
\end{minipage}
\begin{minipage}[c]{.49\textwidth}
 \includegraphics[width=7.5cm]{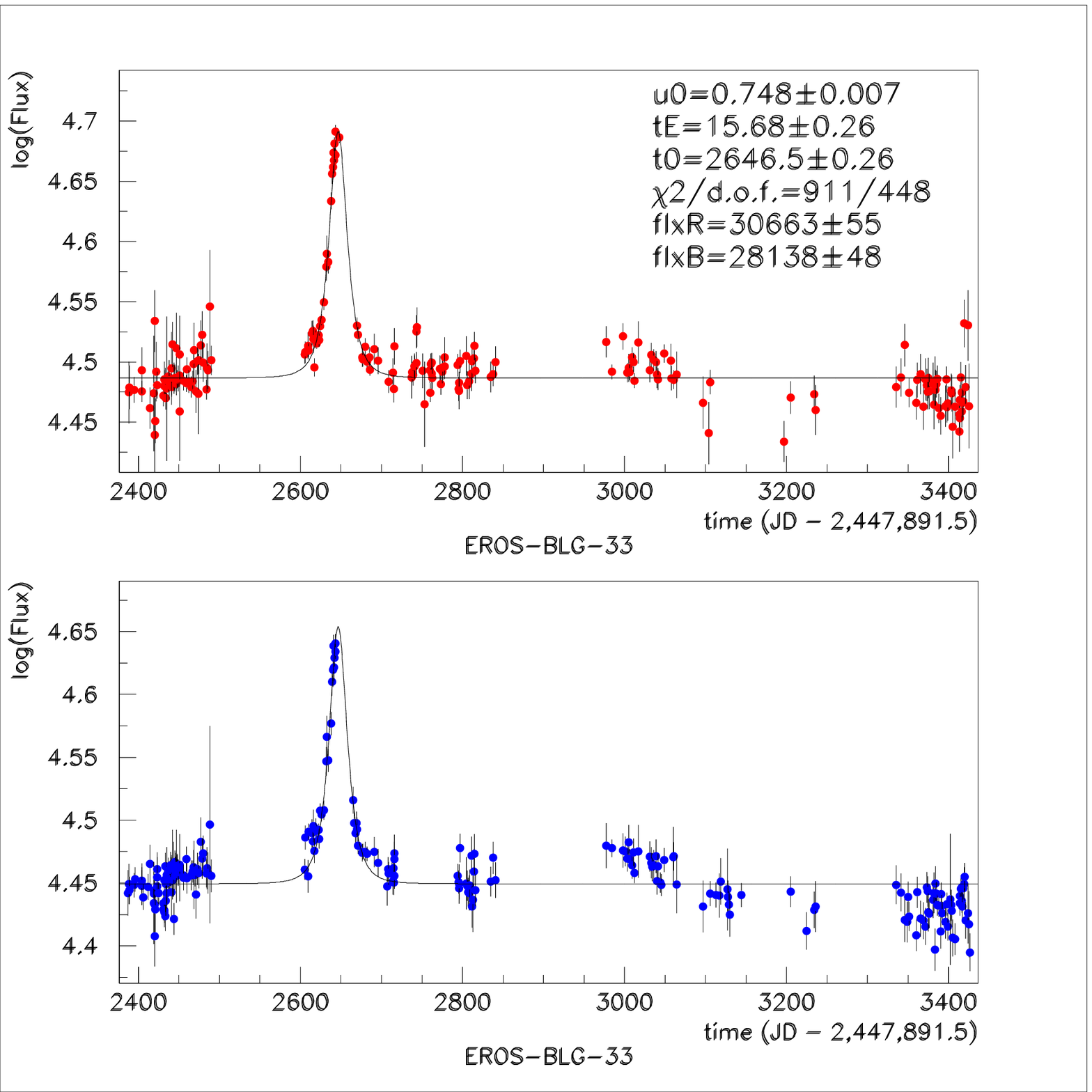}
\end{minipage}
 \begin{minipage}[c]{.49\textwidth}
 \includegraphics[width=7.5cm]{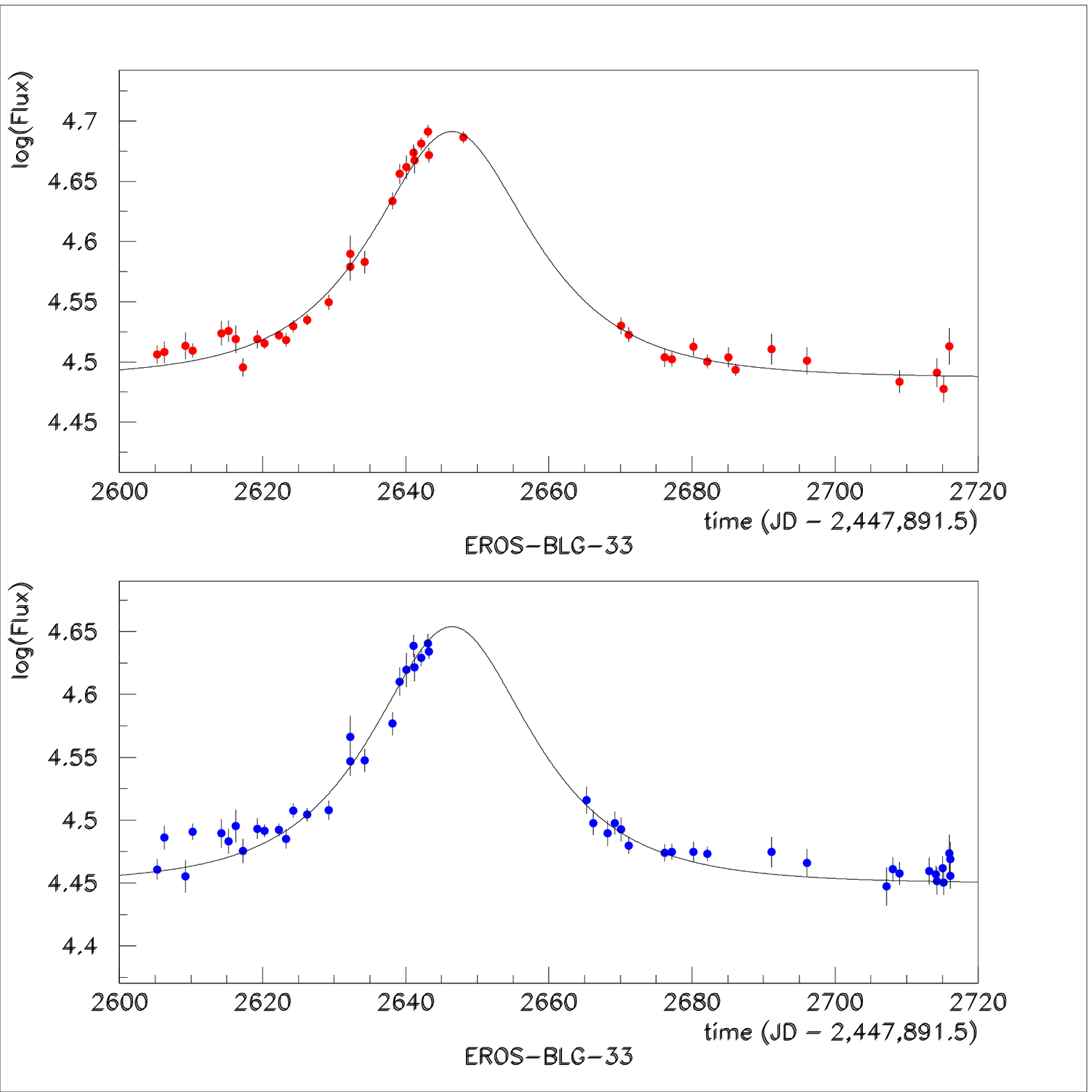}
 \end{minipage}
 \begin{minipage}[c]{.49\textwidth}
 \includegraphics[width=7.5cm]{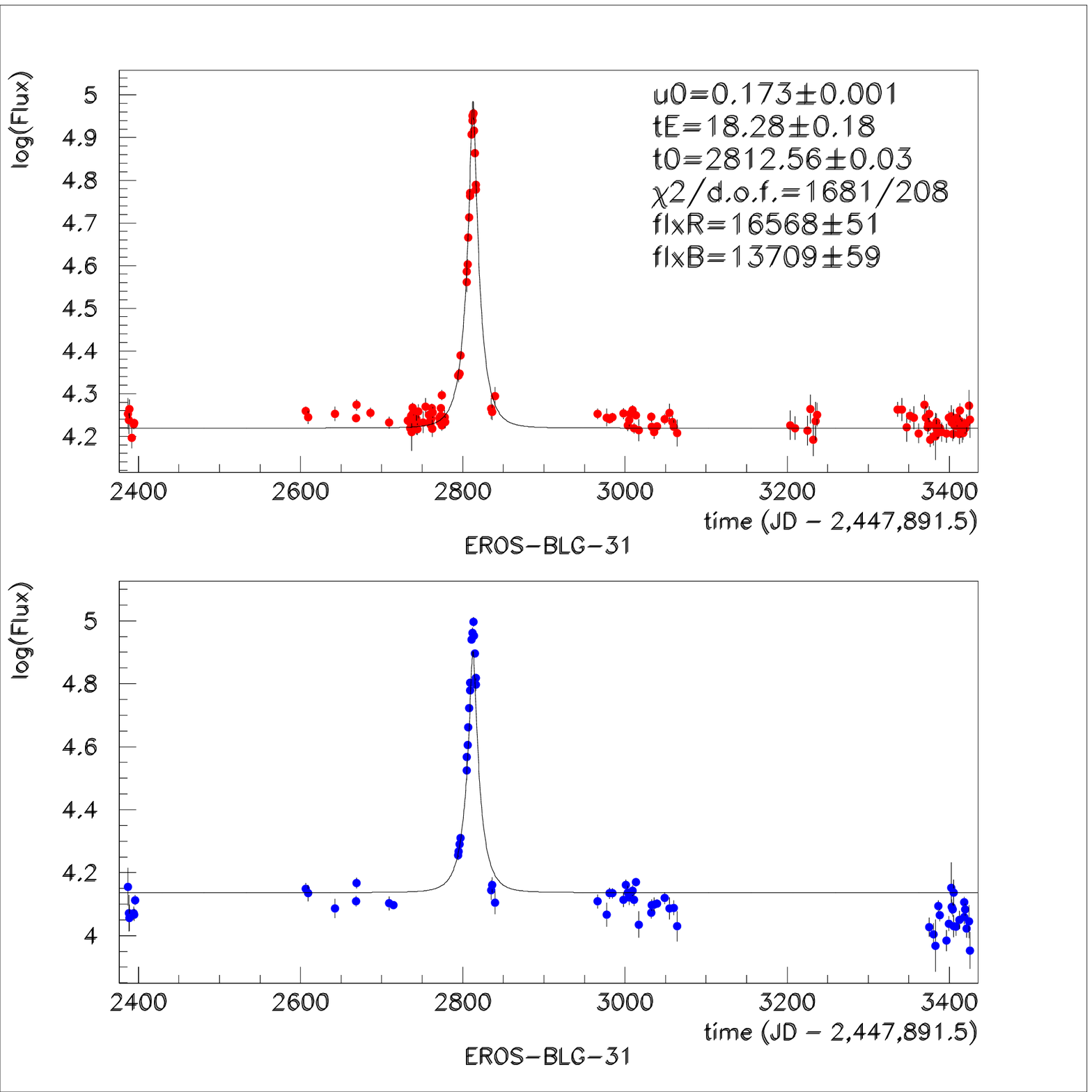}
\end{minipage}
\hspace{0.2cm}
\begin{minipage}[c]{.49\textwidth}
 \includegraphics[width=7.5cm]{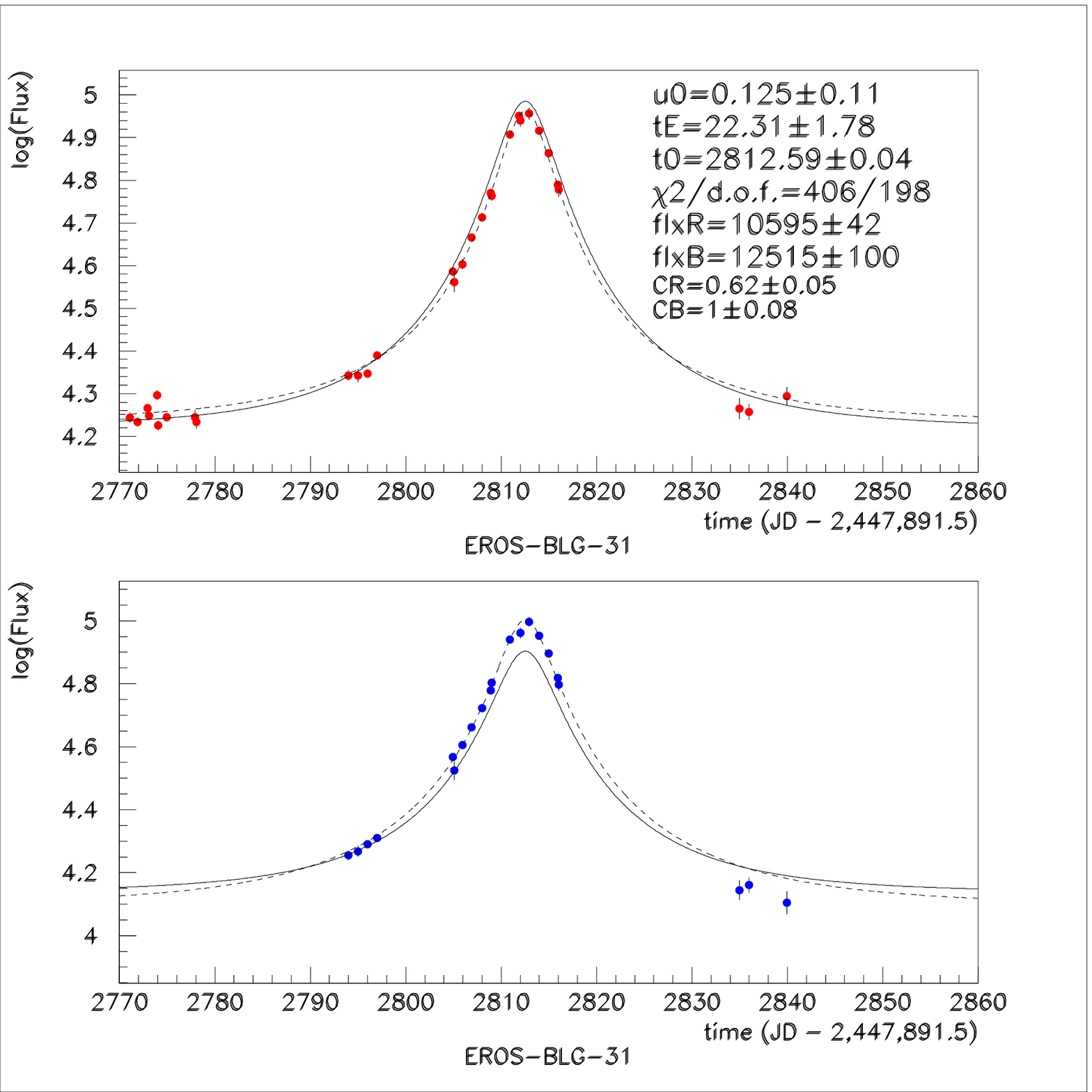}
\end{minipage}
\caption{The light curves of the EROS~2 microlensing candidates
   \#7 to \#9 (see Table~\ref{tab:cand}). In each box the upper
   light curve refers to the EROS red filter and the lower light curve
   to the EROS blue filter. Full span of the light curves is shown in
   the left column and corresponding zoomed light curves are in the
   right column. The 5 parameters obtained by the fit of the
   Paczy\'nski profile are shown (on full span only), as well as the
   $\chi^2$ values of the fit. For candidate \#9 the dashed
   line refers to the fit when blending is taken into account. The
   left light curves of this candidate indicate the parameters of the
   microlensing fit without blending and the zoom (right light curve)
   shows the parameters of the fit with blending.}   
  \label{fig:cdl_candidates2}
\end{figure}

\begin{figure}[h]
 \begin{minipage}[c]{.49\textwidth}
 \includegraphics[width=7.5cm]{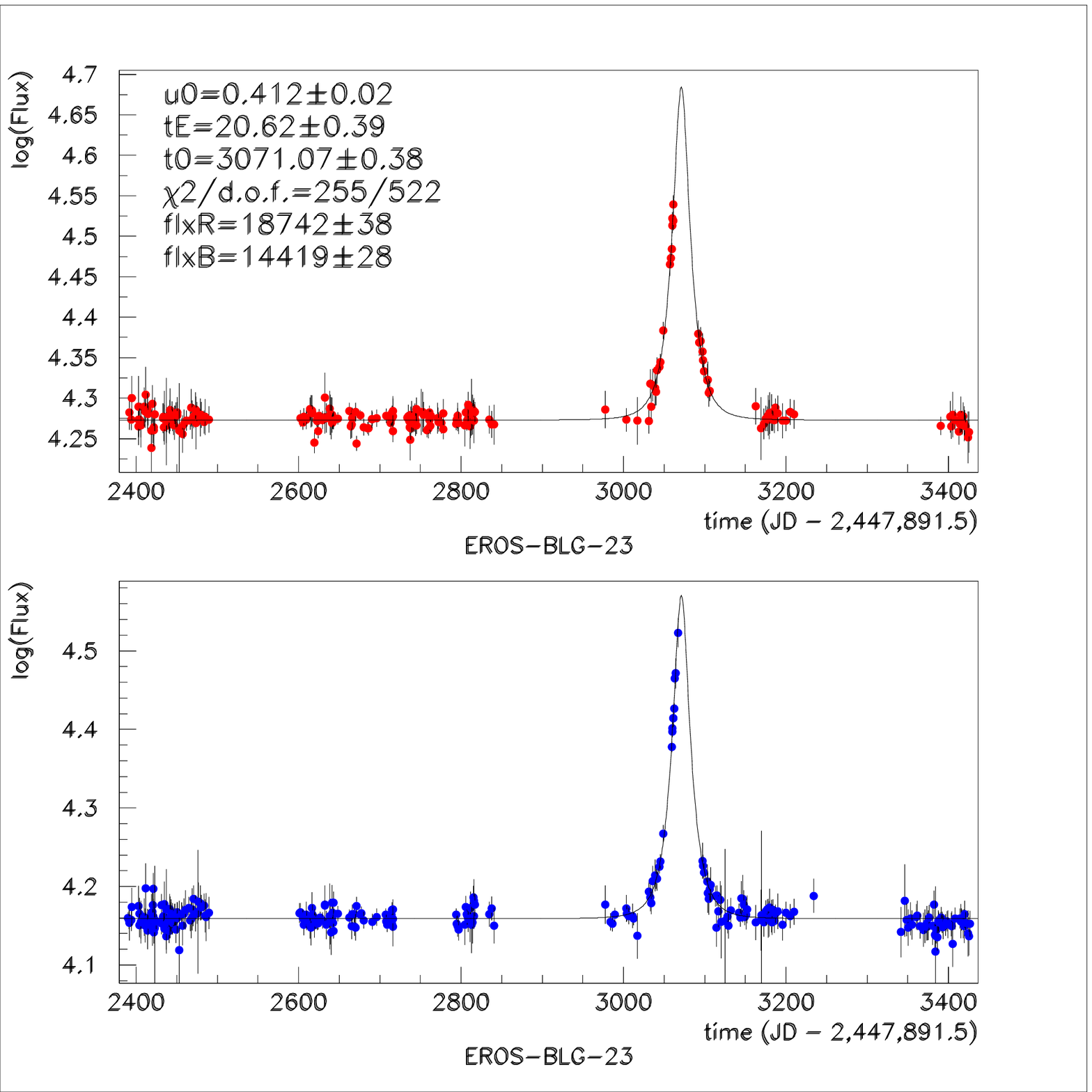}
 \end{minipage}
 \begin{minipage}[c]{.49\textwidth}
 \includegraphics[width=7.5cm]{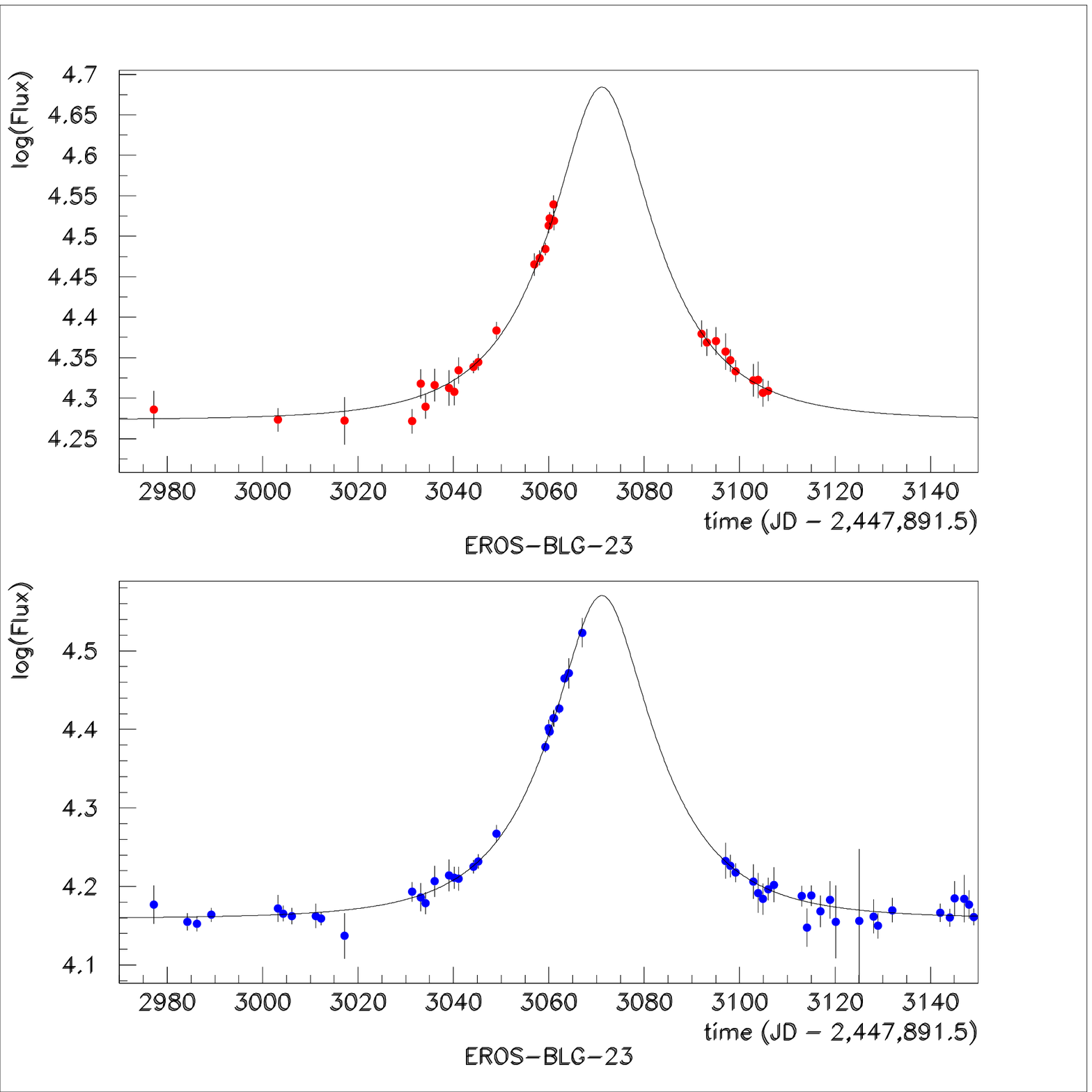}
\end{minipage}
\begin{minipage}[c]{.49\textwidth}
 \includegraphics[width=7.5cm]{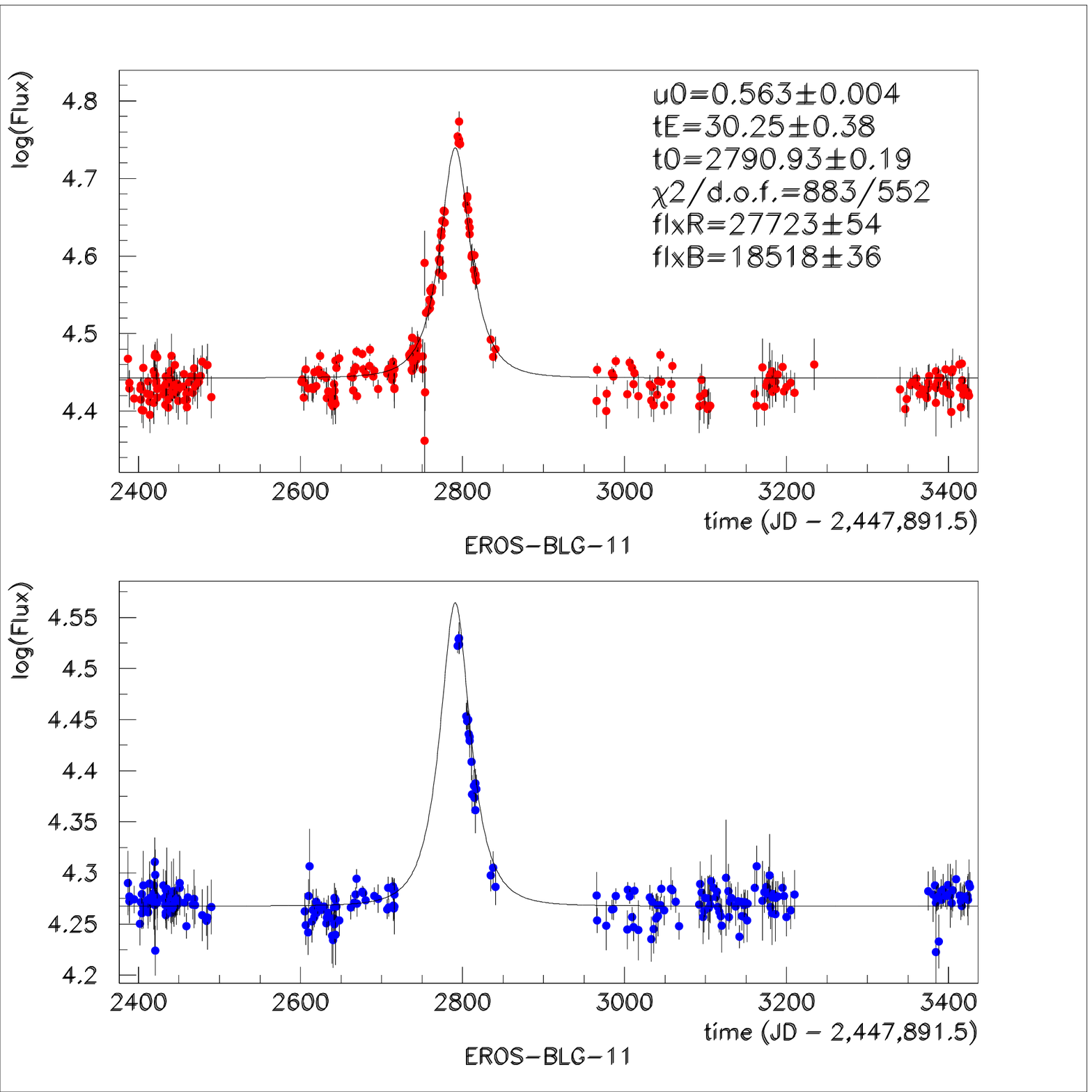}
\end{minipage}
 \begin{minipage}[c]{.49\textwidth}
 \includegraphics[width=7.5cm]{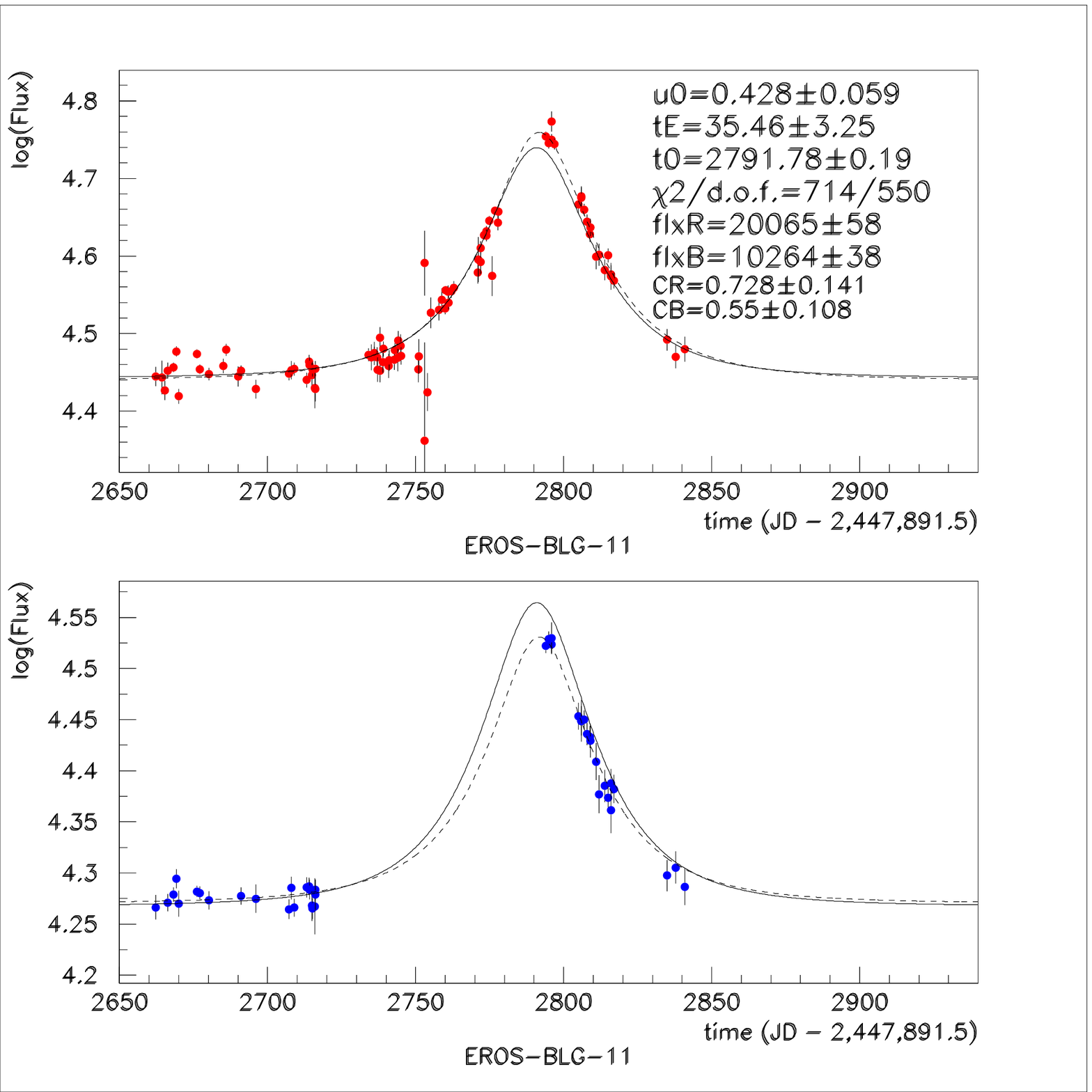}
 \end{minipage}
 \begin{minipage}[c]{.49\textwidth}
 \includegraphics[width=7.5cm]{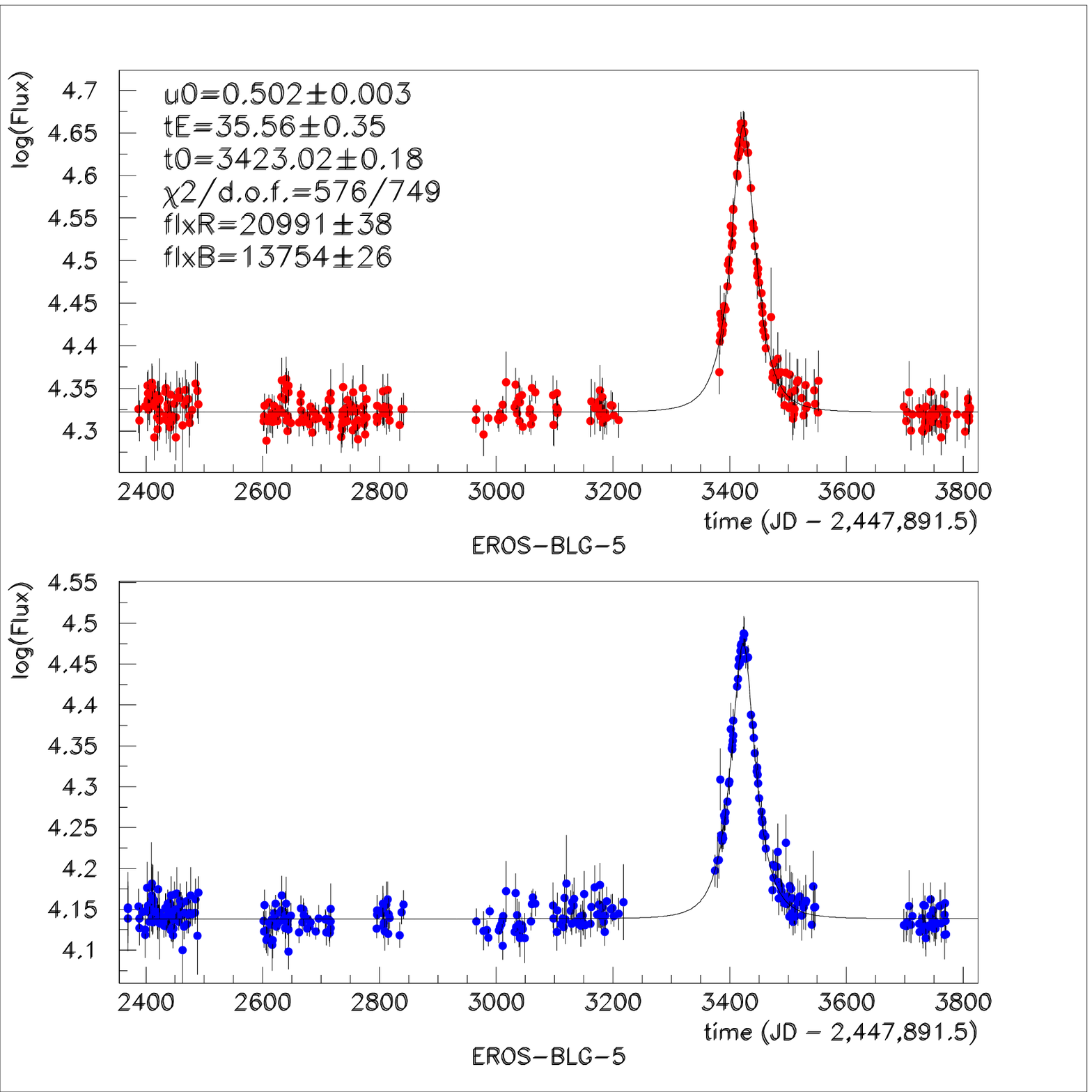}
 \end{minipage}
\hspace{0.2cm}
 \begin{minipage}[c]{.49\textwidth}
 \includegraphics[width=7.5cm]{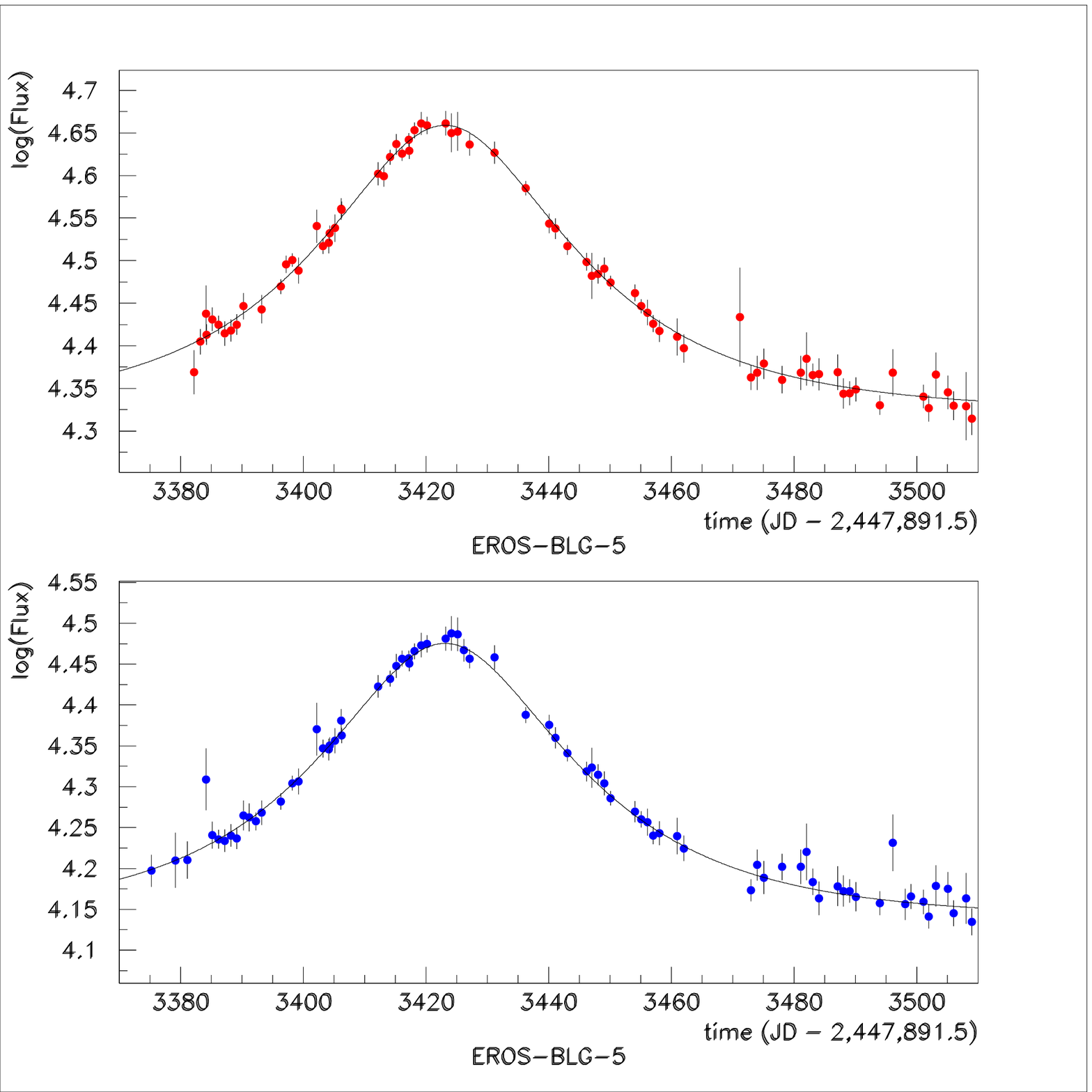}
 \end{minipage}
 \caption{The light curves of the EROS~2 microlensing candidates
   \#10 to \#12 (see Table~\ref{tab:cand}). In each box the upper
   light curve refers to the EROS red filter and the lower light curve
   to the EROS blue filter. Full span of the light curves is shown in
   the left column and corresponding zoomed light curves are in the
   right column. The 5 parameters obtained by the fit of the
   Paczy\'nski profile are shown (on full span), as well as the $\chi^2$ 
   values of the fit. For candidate \#11 the dashed
   line refers to the fit when blending is taken into account. The
   left light curves of this candidate indicate the parameters of the
   microlensing fit without blending and the zoom (right light curve)
   shows the parameters of the fit with blending.}  
  \label{fig:cdl_candidates3}
\end{figure}

\begin{figure}[h]
 \begin{minipage}[c]{.49\textwidth}
 \includegraphics[width=7.5cm]{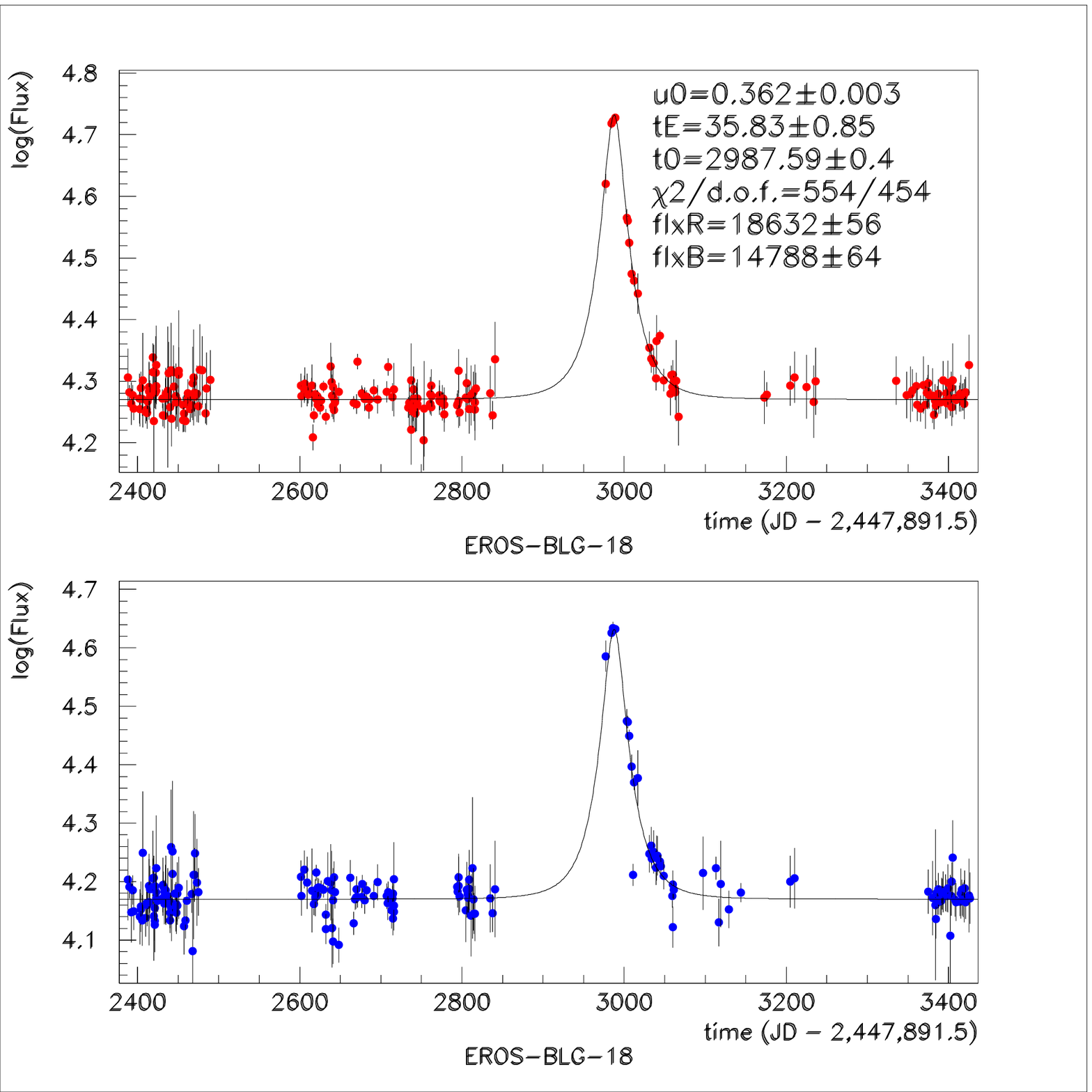}
 \end{minipage}
 \begin{minipage}[c]{.49\textwidth}
 \includegraphics[width=7.5cm]{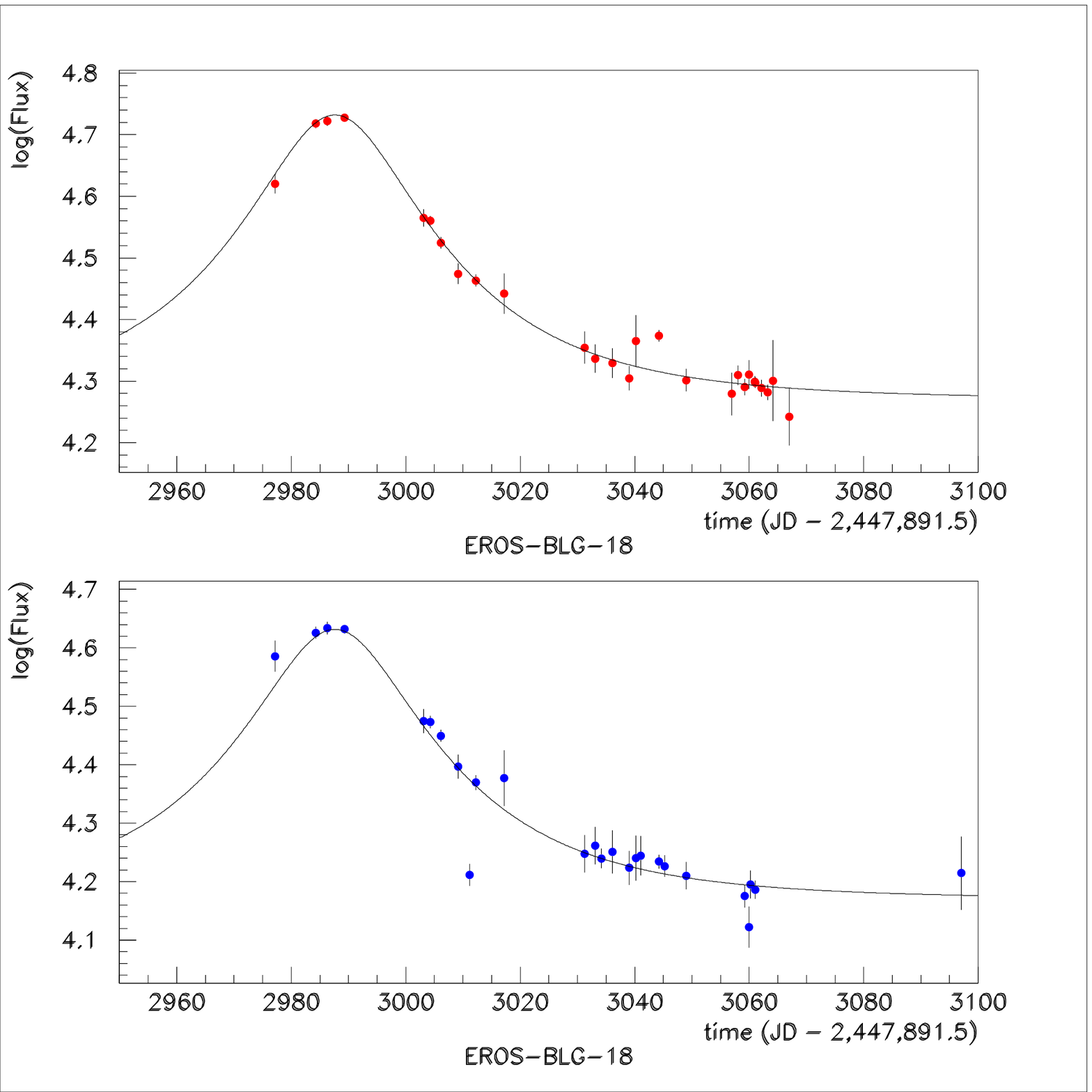}
\end{minipage}
\begin{minipage}[c]{.49\textwidth}
 \includegraphics[width=7.5cm]{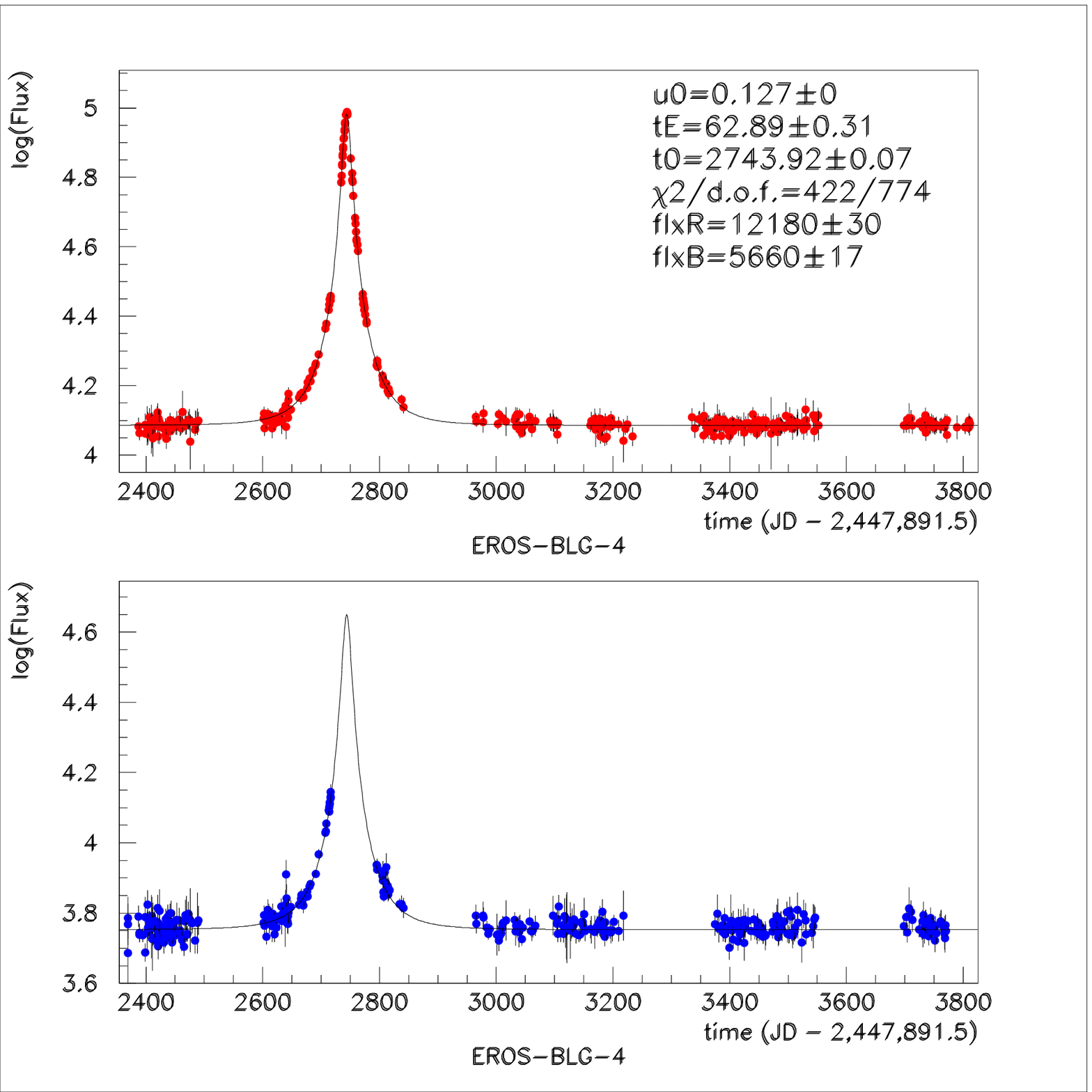}
\end{minipage}
 \begin{minipage}[c]{.49\textwidth}
 \includegraphics[width=7.5cm]{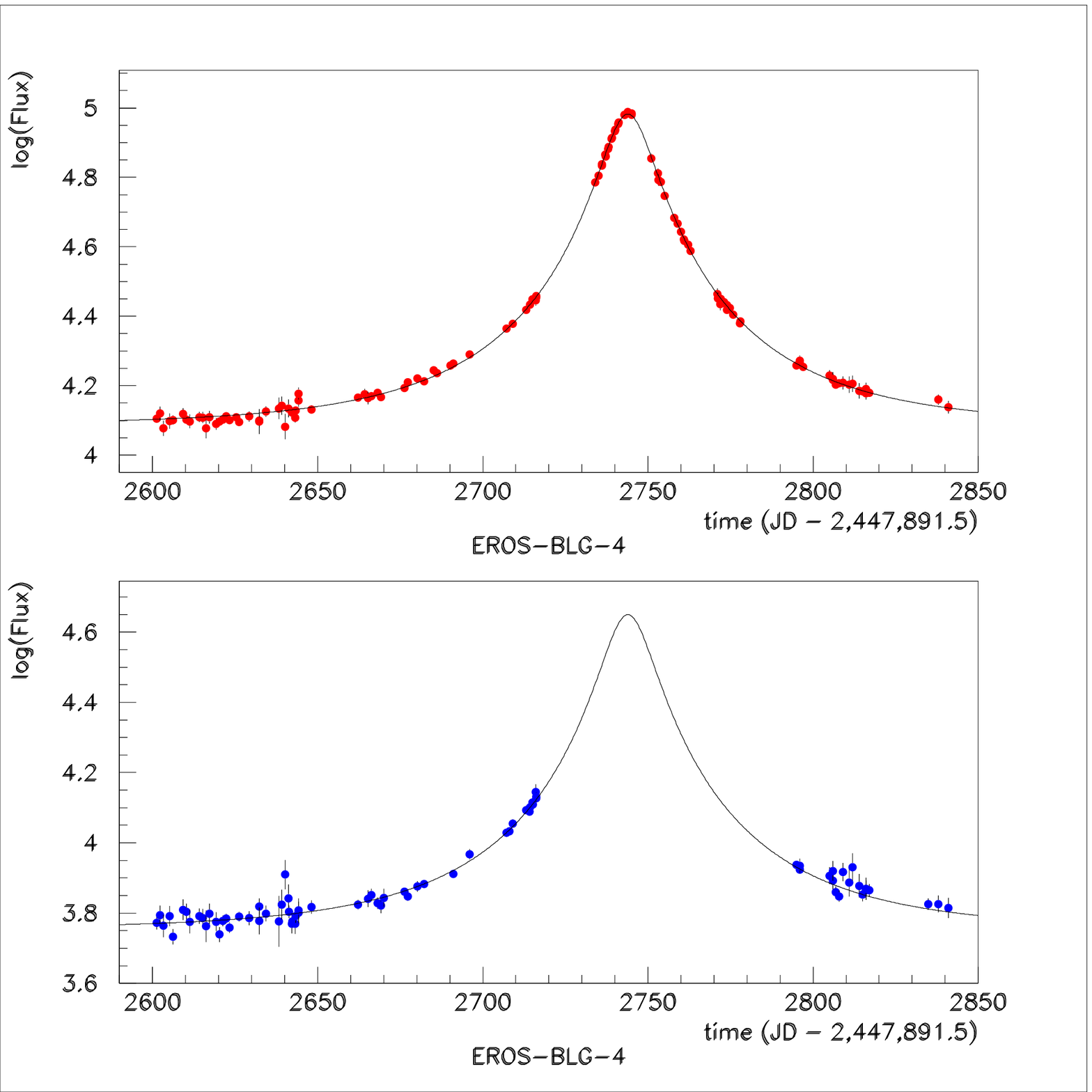}
 \end{minipage}
 \begin{minipage}[c]{.49\textwidth}
 \includegraphics[width=7.5cm]{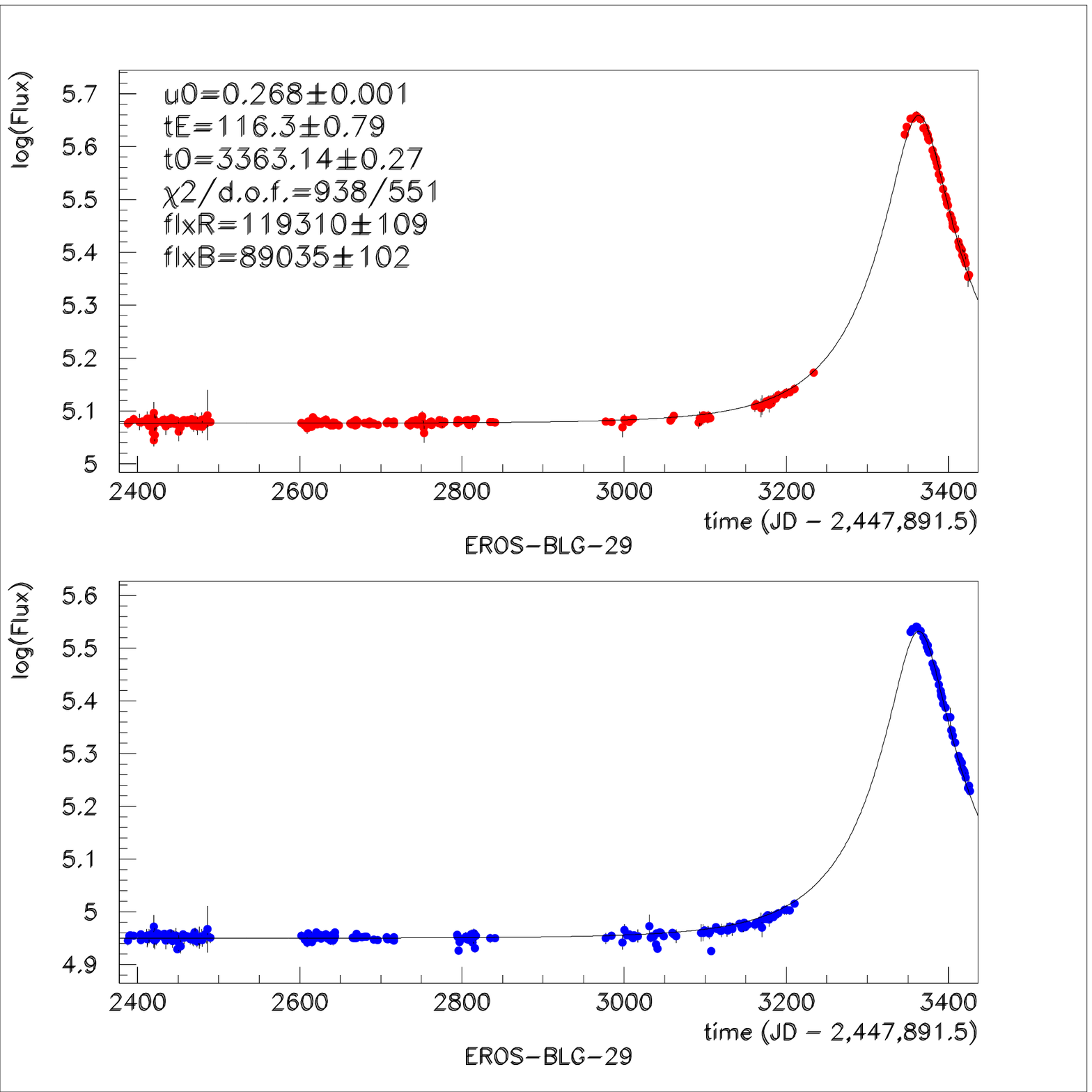}
 \end{minipage}
\hspace{0.2cm}
 \begin{minipage}[c]{.49\textwidth}
 \includegraphics[width=7.5cm]{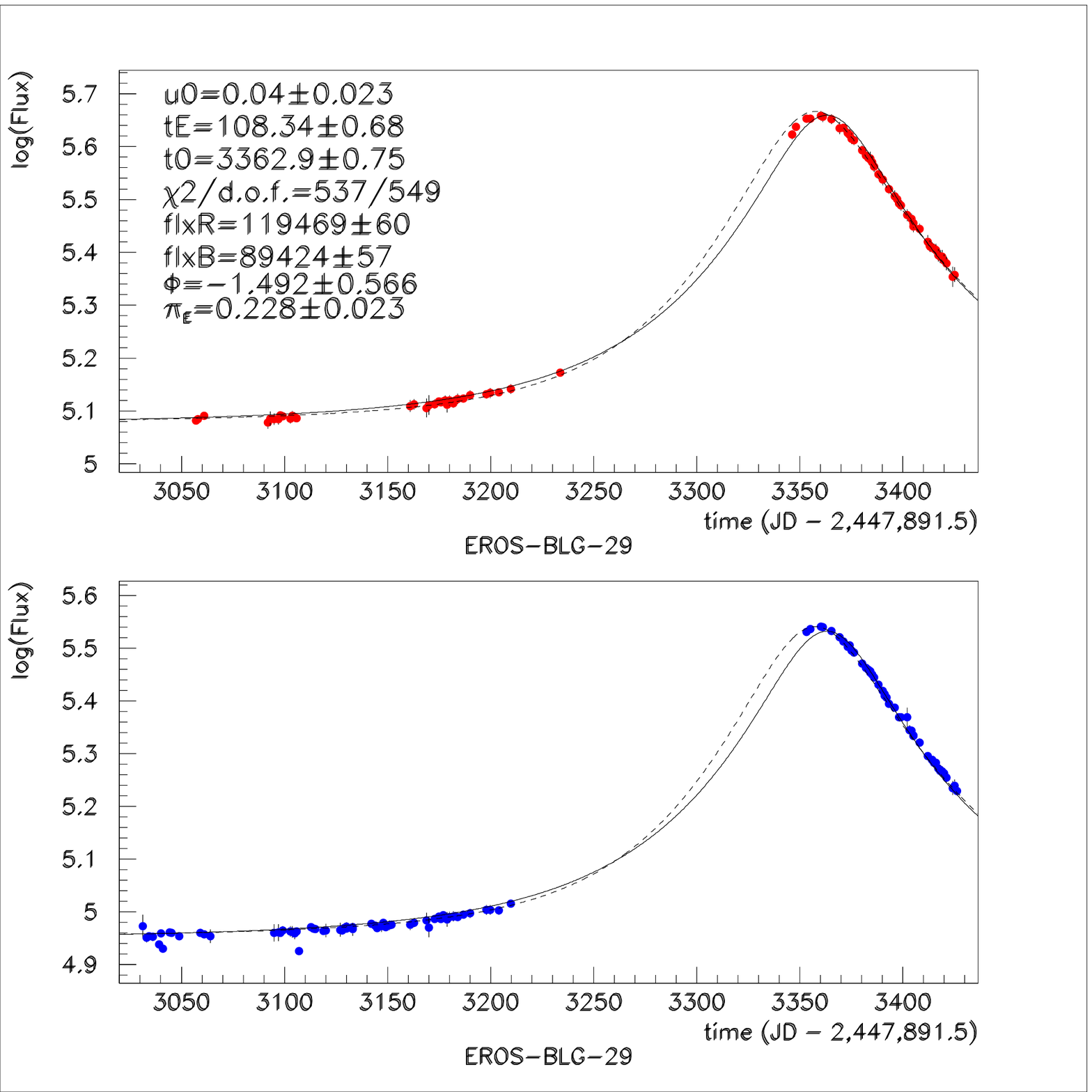}
 \end{minipage}
 \caption{The light curves of the EROS~2 microlensing candidates
   \#13 to \#15 (see Table~\ref{tab:cand}). In each box the upper
   light curve refers to the EROS red filter and the lower light curve
   to the EROS blue filter. Full span of the light curves is shown in
   the left column and corresponding zoomed light curves are in the
   right column. The 5 parameters obtained by the
   fit of the Paczy\'nski profile are shown (on full span), as well as
   the $\chi^2$ values of the fit. For candidate \#15 the dashed
   line refers to the fit when parallax is taken into account. The
   left light curves of this candidate indicate the parameters of the
   microlensing fit without parallax and the zoom (right light curves)
   shows the parameters of the fit with parallax.}  
  \label{fig:cdl_candidates4}
\end{figure}

\begin{figure}[h]
 \begin{minipage}[c]{.49\textwidth}
 \includegraphics[width=7.5cm]{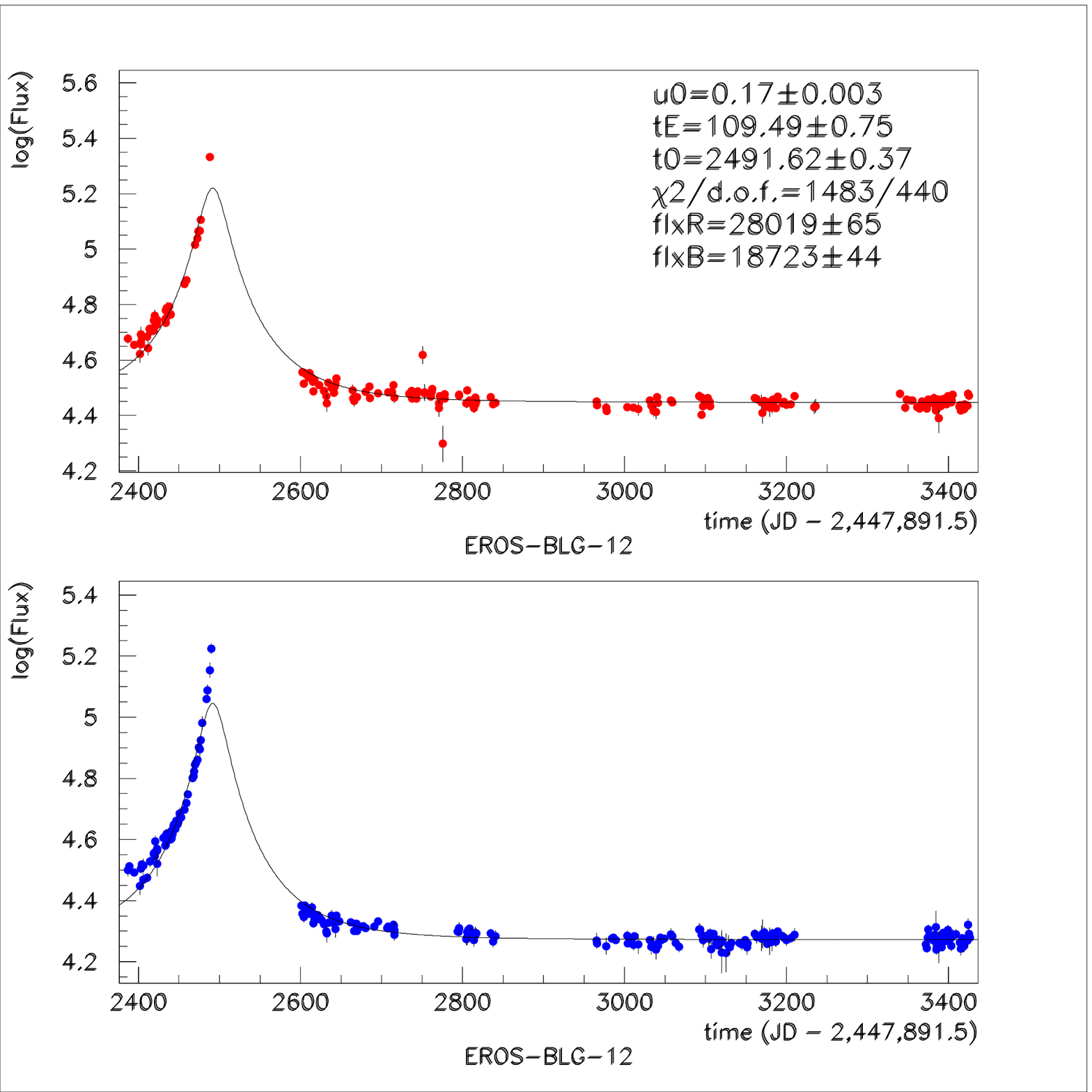}
 \end{minipage}
 \begin{minipage}[c]{.49\textwidth}
 \includegraphics[width=7.5cm]{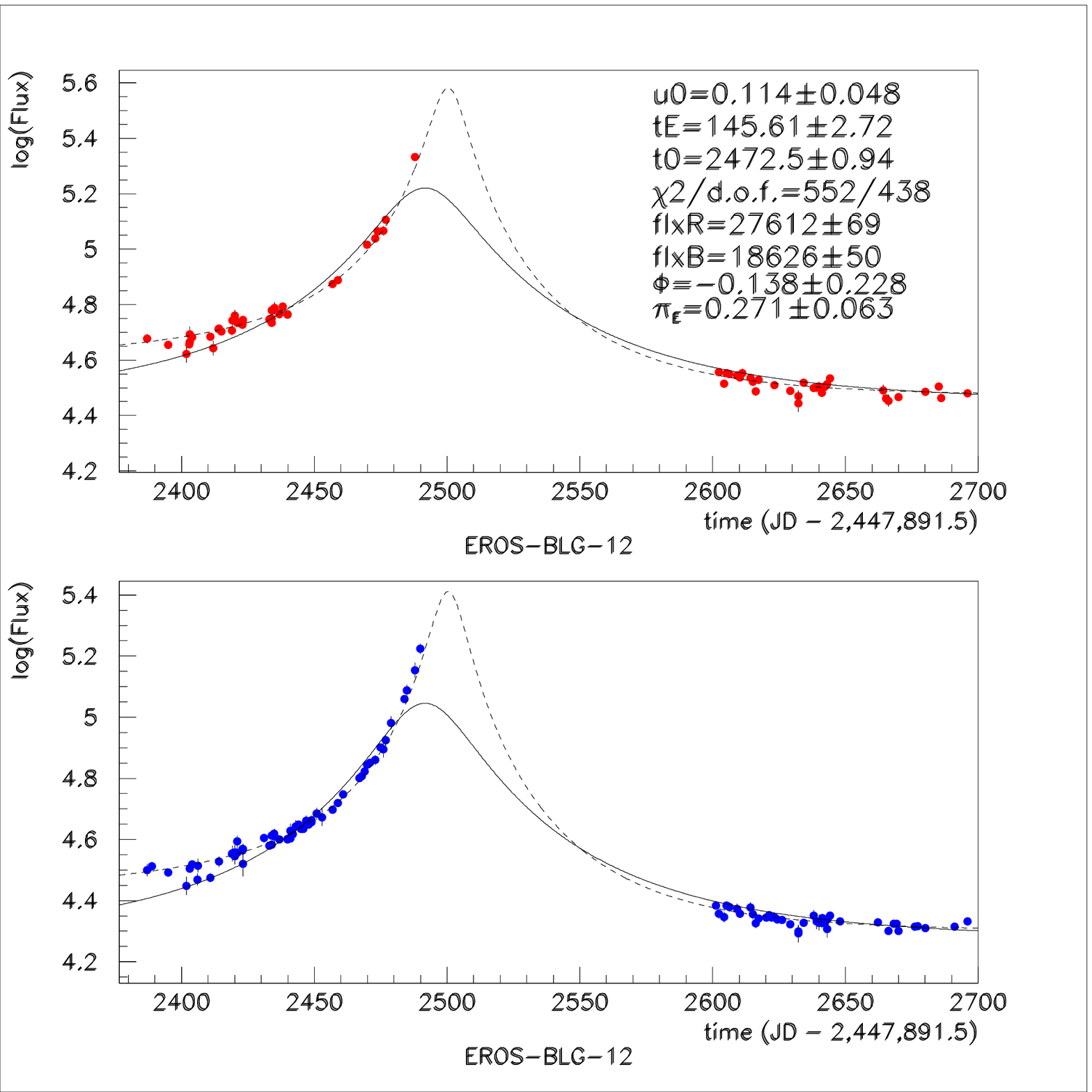}
\end{minipage}
 \caption{The light curves of the EROS~2 microlensing candidate
   \#16 (see Table~\ref{tab:cand}). In each box the upper
   light curve refers to the EROS red filter and the lower light curve
   to the EROS blue filter. Full span of the light curves is shown in
   the left column and corresponding zoomed light curves are in the
   right column. The 5 parameters obtained by the
   fit of the Paczy\'nski profile are shown (on full span), as well as
   the $\chi^2$ values of the fit. For candidate \#16 the dashed line
   refers to the fit when parallax is taken into account. The left
   light curves of this candidate indicate the parameters of the
   microlensing fit without parallax and the zoom (right light curves)
   shows the parameters of the fit with parallax.}   
  \label{fig:cdl_candidates5}
\end{figure}

    To obtain a reliable value for the bulge optical depth to
microlensing, the least affected by systematic errors due to
blending, we decided to consider only events with clump-giant 
sources (of the extended clump area), and to make a final cut
requiring $u_0<1$, because it is 
difficult to totally rule out other forms of stellar variability for
lower amplification events. This selection yielded 16 events. In 2
cases, we found that the microlensing fit was improved
by adding two additional parameters for parallax (\cite{GOU92}),
$\pi_{\rm E}$, the amplitude of the displacement in the Einstein ring due to
the Earth's orbital motion, and $\phi$, the phase of that
displacement (see Table~\ref{tab:cand}, Fig.~\ref{fig:cdl_candidates4} and
Fig.~\ref{fig:cdl_candidates5}). We also searched for blending effects
on the selected sample of 16 candidates. Two light curves seem to be affected,
showing a significant improvement of the microlensing fit when
blending is taken into account, particularly for candidate \#9 (see
Table~\ref{tab:cand}, Fig.~\ref{fig:cdl_candidates2} and
Fig.~\ref{fig:cdl_candidates3}).

\begin{figure}[h]
 \centering
 \includegraphics[width=7.8cm]{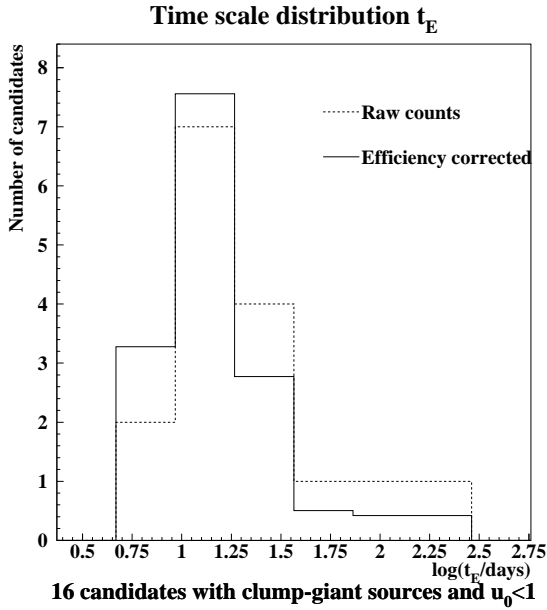}
 \caption{Time scales distribution of the 16 microlensing candidates
   with clump-giant sources (of the extended clump area) and
   $u_0<1$ $(A_0>1.34)$. The dashed line shows the 
   raw data, while the solid curve is corrected for the detection
   efficiency. For the sake of comparison, the distribution of the
   corrected data was scaled so that the two histograms have the same area.}
  \label{fig:distri_evt_te}
\end{figure}

\begin{figure}[h]
\begin{minipage}[c]{.49\textwidth}
\includegraphics[width=7.5cm]{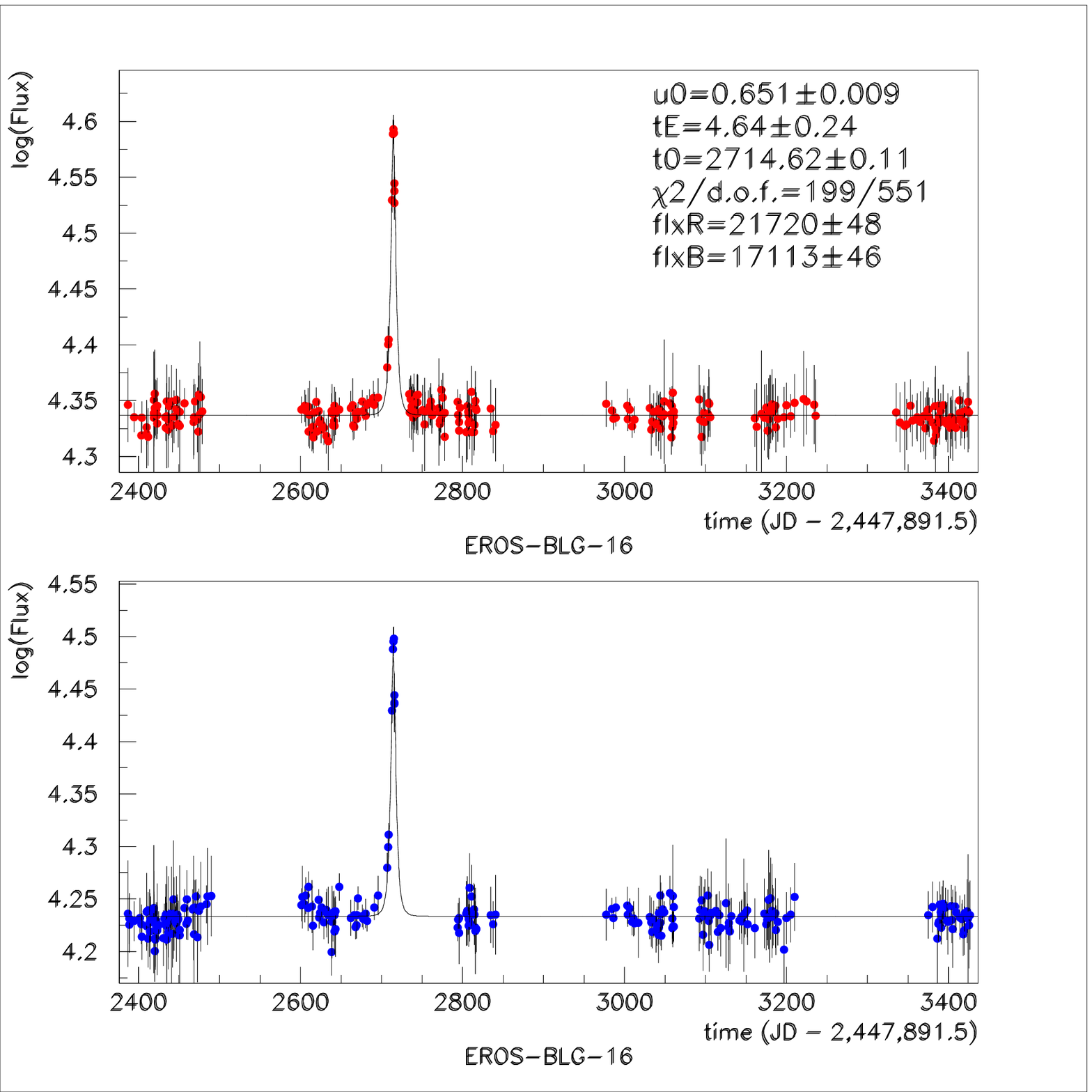}
\end{minipage}
\begin{minipage}[c]{.49\textwidth}
\includegraphics[width=7.5cm]{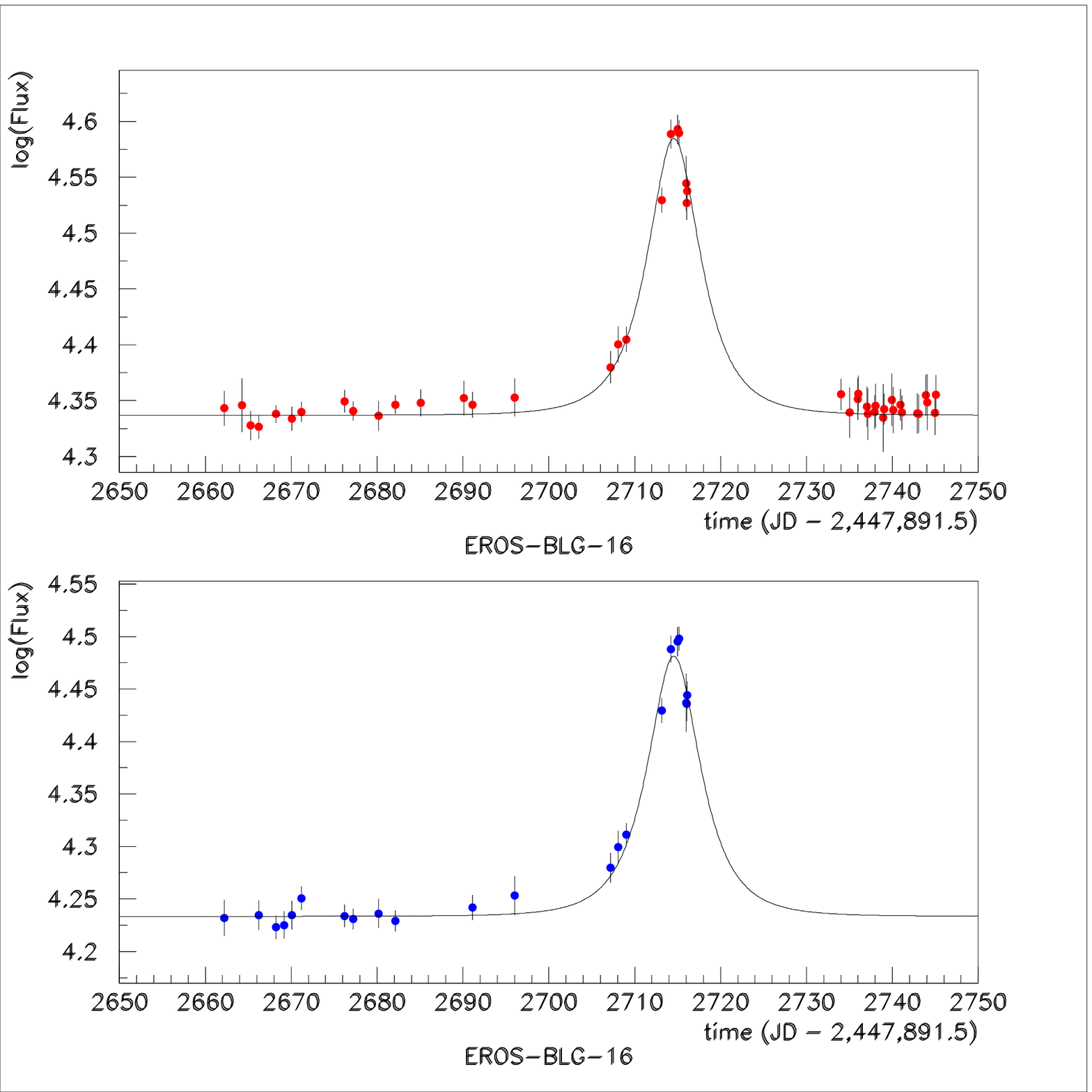}
\end{minipage}
\begin{minipage}[c]{.49\textwidth}
 \includegraphics[width=7.5cm]{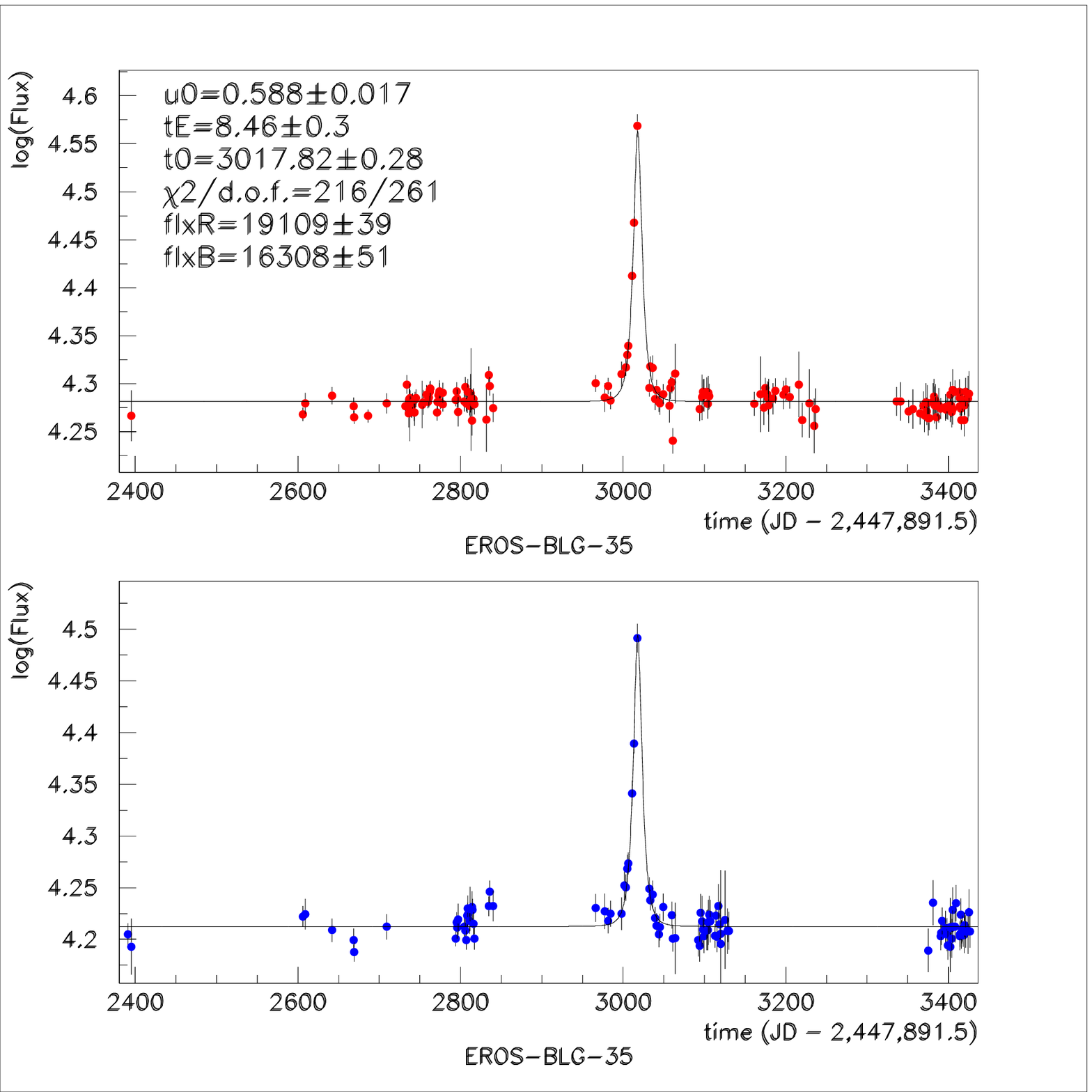}
\end{minipage}
 \begin{minipage}[c]{.49\textwidth}
 \includegraphics[width=7.5cm]{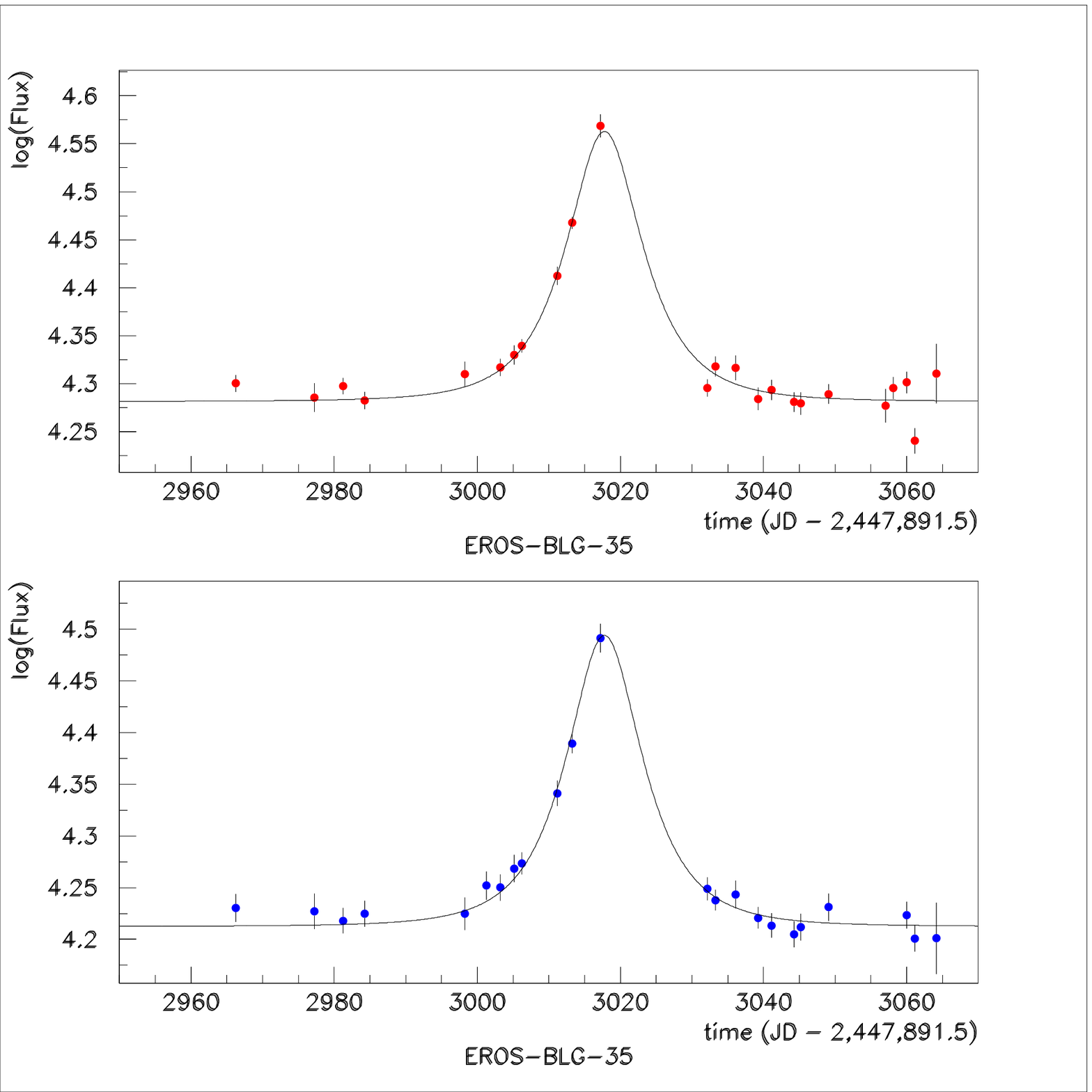}
 \end{minipage}
 \begin{minipage}[c]{.49\textwidth}
 \includegraphics[width=7.5cm]{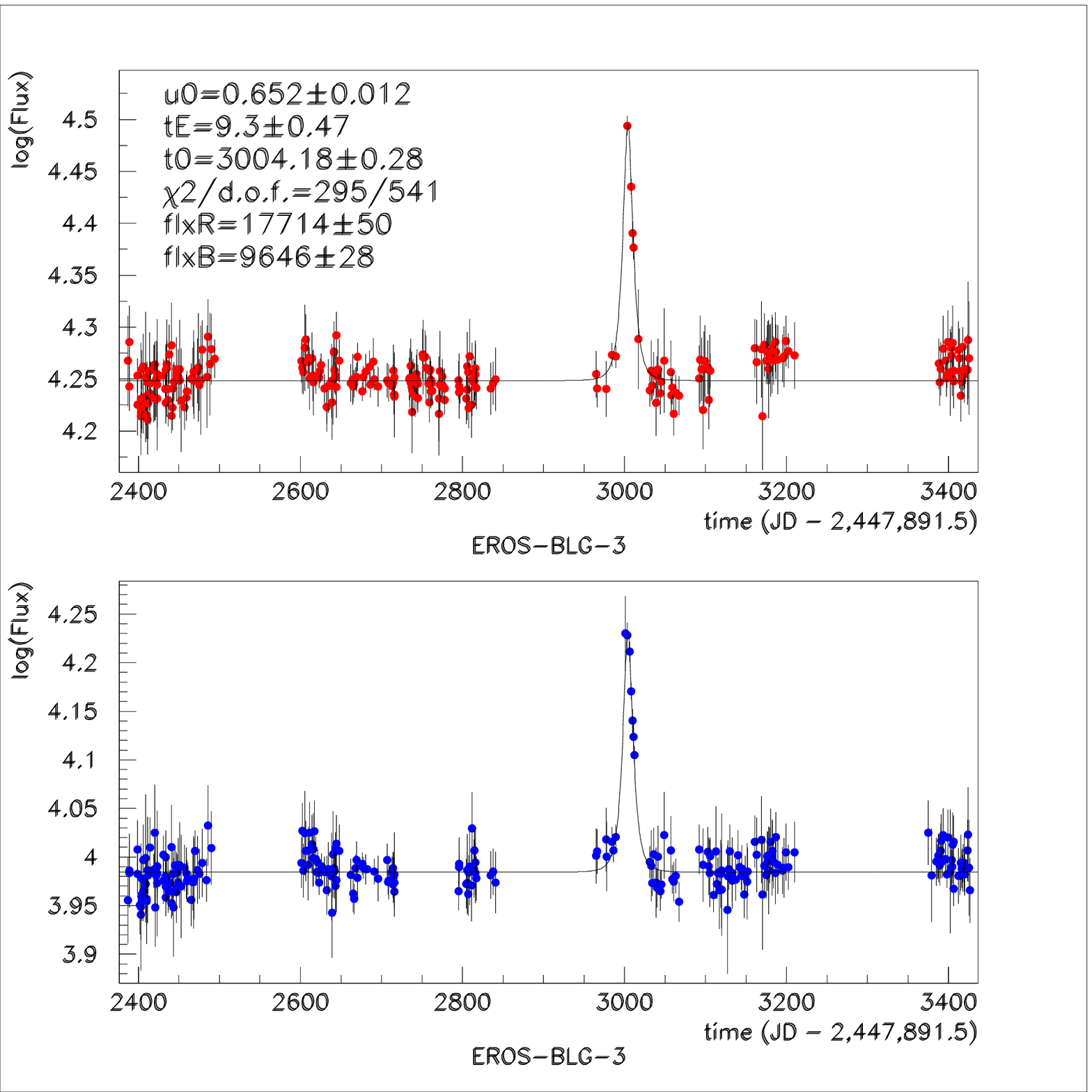}
\end{minipage}
\hspace{0.2cm}
\begin{minipage}[c]{.49\textwidth}
\includegraphics[width=7.5cm]{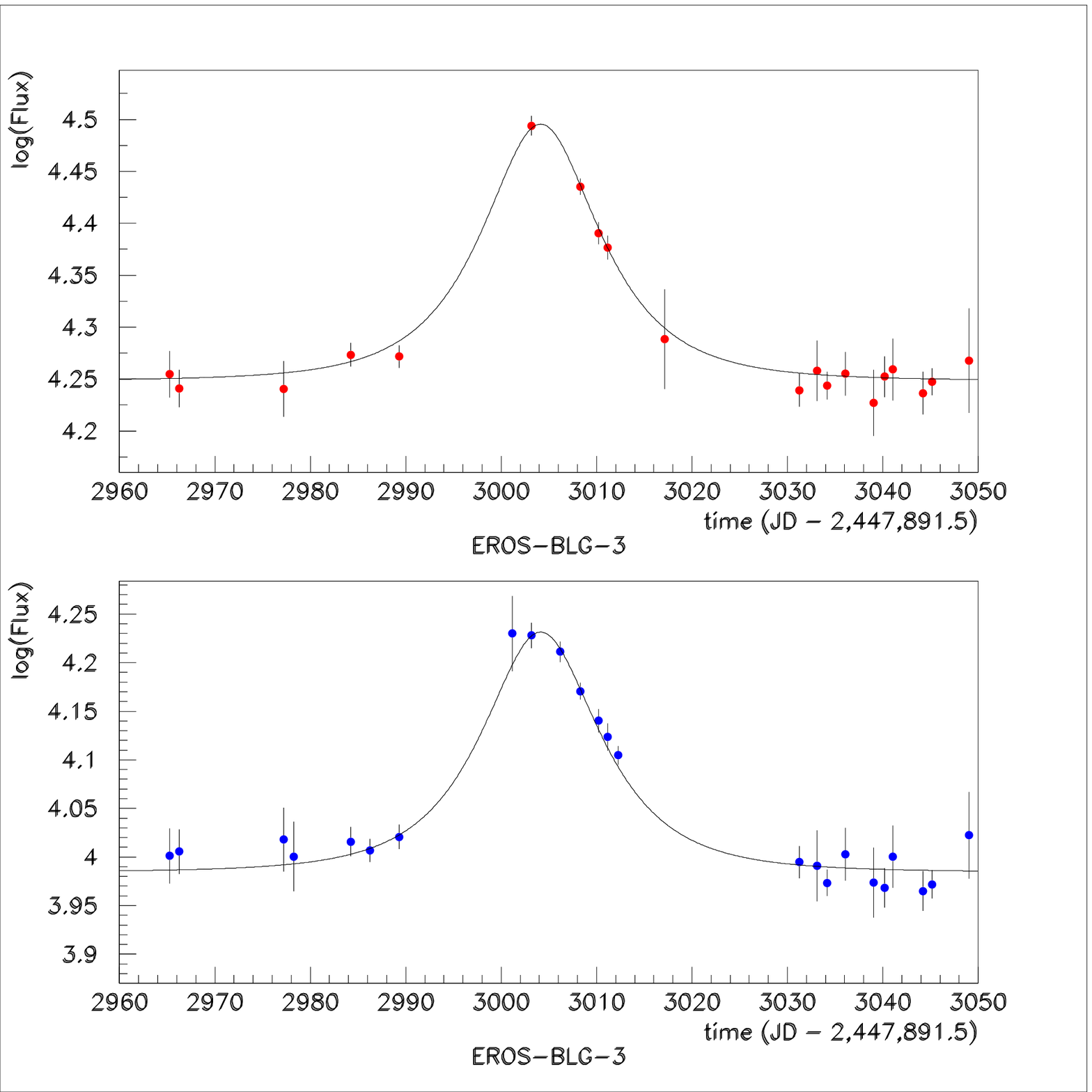}
\end{minipage}
\caption{The light curves of the EROS~2 microlensing candidates
   \#1 to \#3 (see Table~\ref{tab:cand}). In each box the upper
   light curve refers to the EROS red filter and the lower light curve
   to the EROS blue filter. Full span of the light curves is shown in
   the left column and corresponding zoomed light curves are in the
   right column. The 5 parameters obtained by the fit of the
   Paczy\'nski profile are shown (on full span only), as well as the
   $\chi^2$ values of the fit.}  
  \label{fig:cdl_candidates}
\end{figure}

\begin{figure}[h]
 \begin{minipage}[c]{.49\textwidth}
 \includegraphics[width=7.5cm]{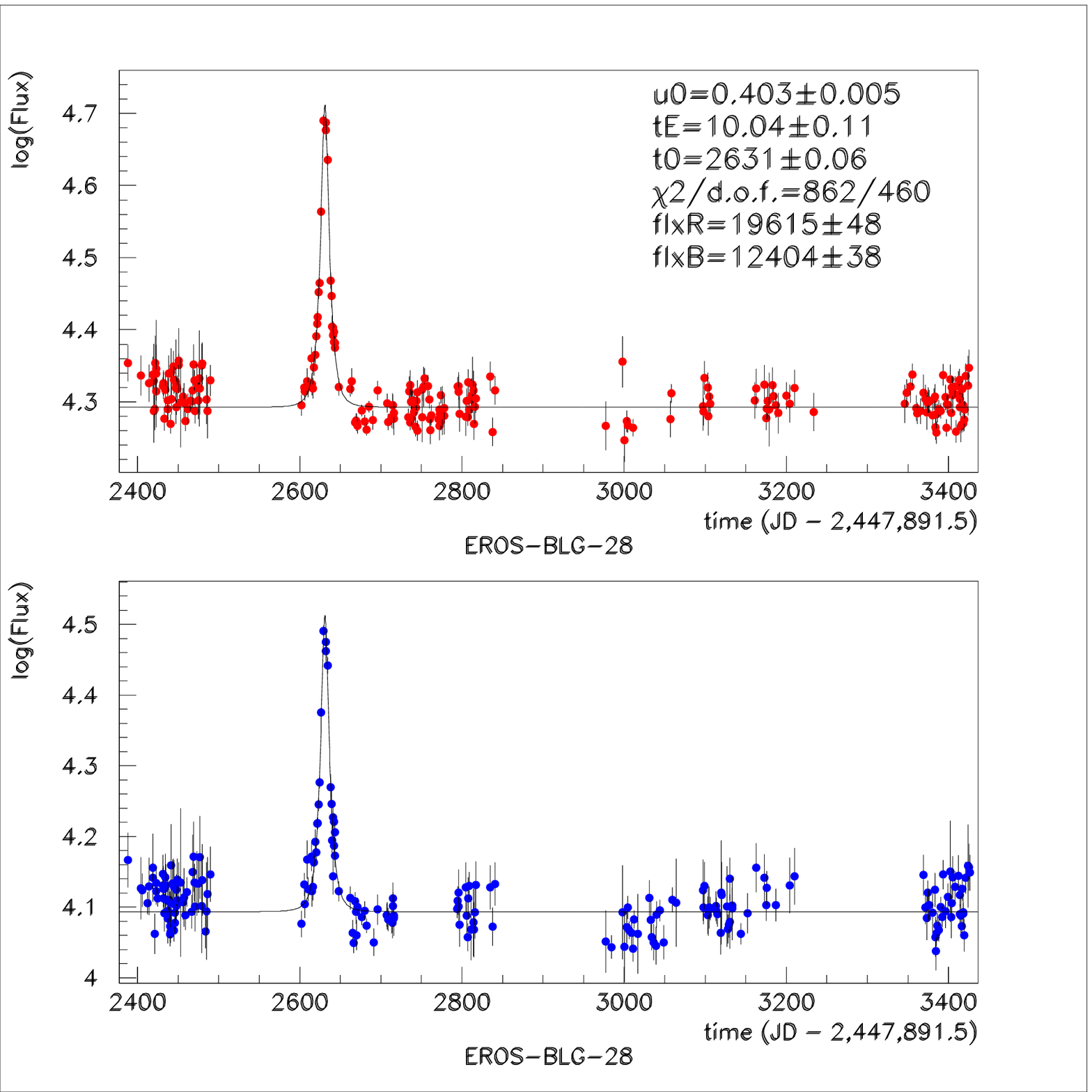}
 \end{minipage}
 \begin{minipage}[c]{.49\textwidth}
 \includegraphics[width=7.5cm]{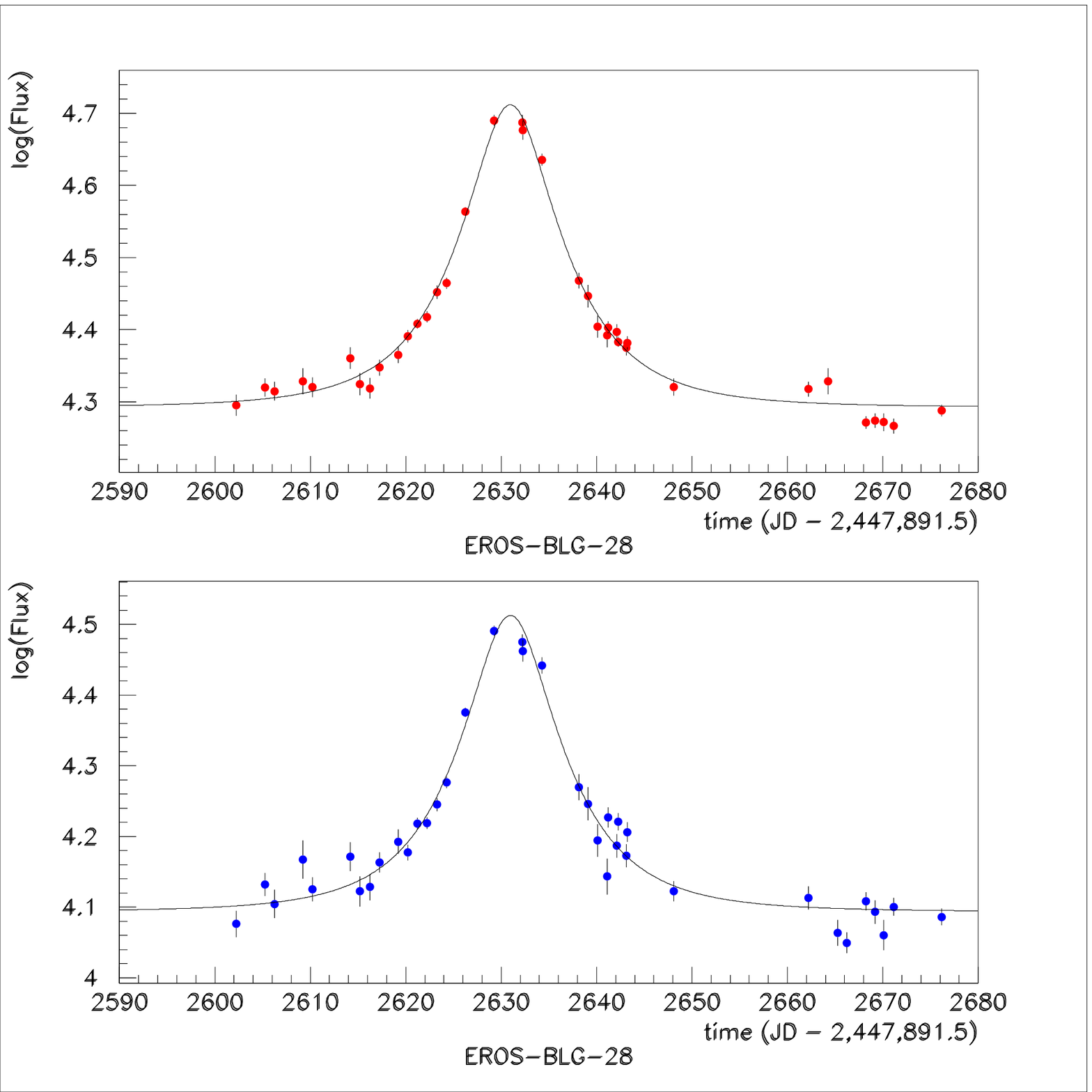}
\end{minipage}
\begin{minipage}[c]{.49\textwidth}
 \includegraphics[width=7.5cm]{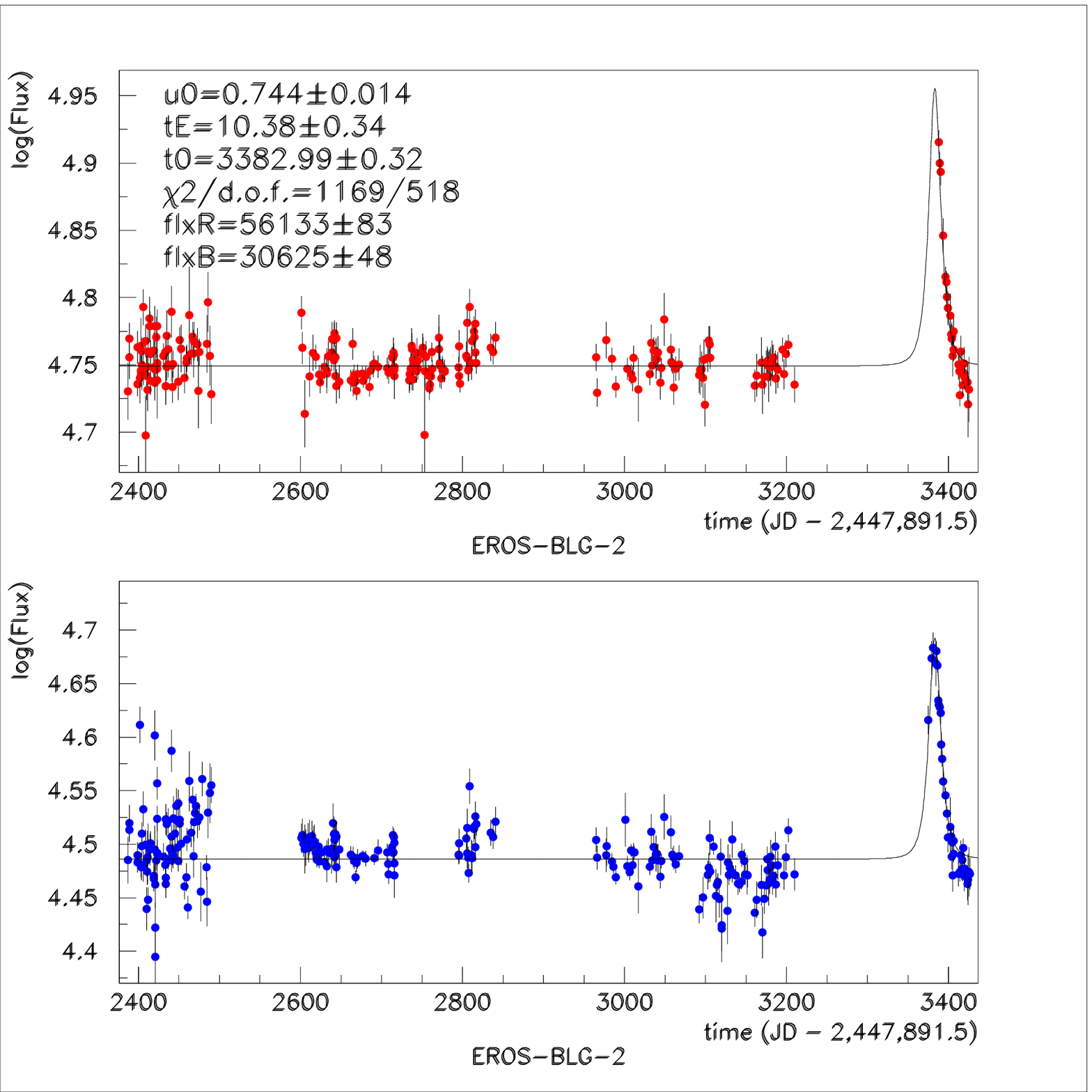}
\end{minipage}
 \begin{minipage}[c]{.49\textwidth}
 \includegraphics[width=7.5cm]{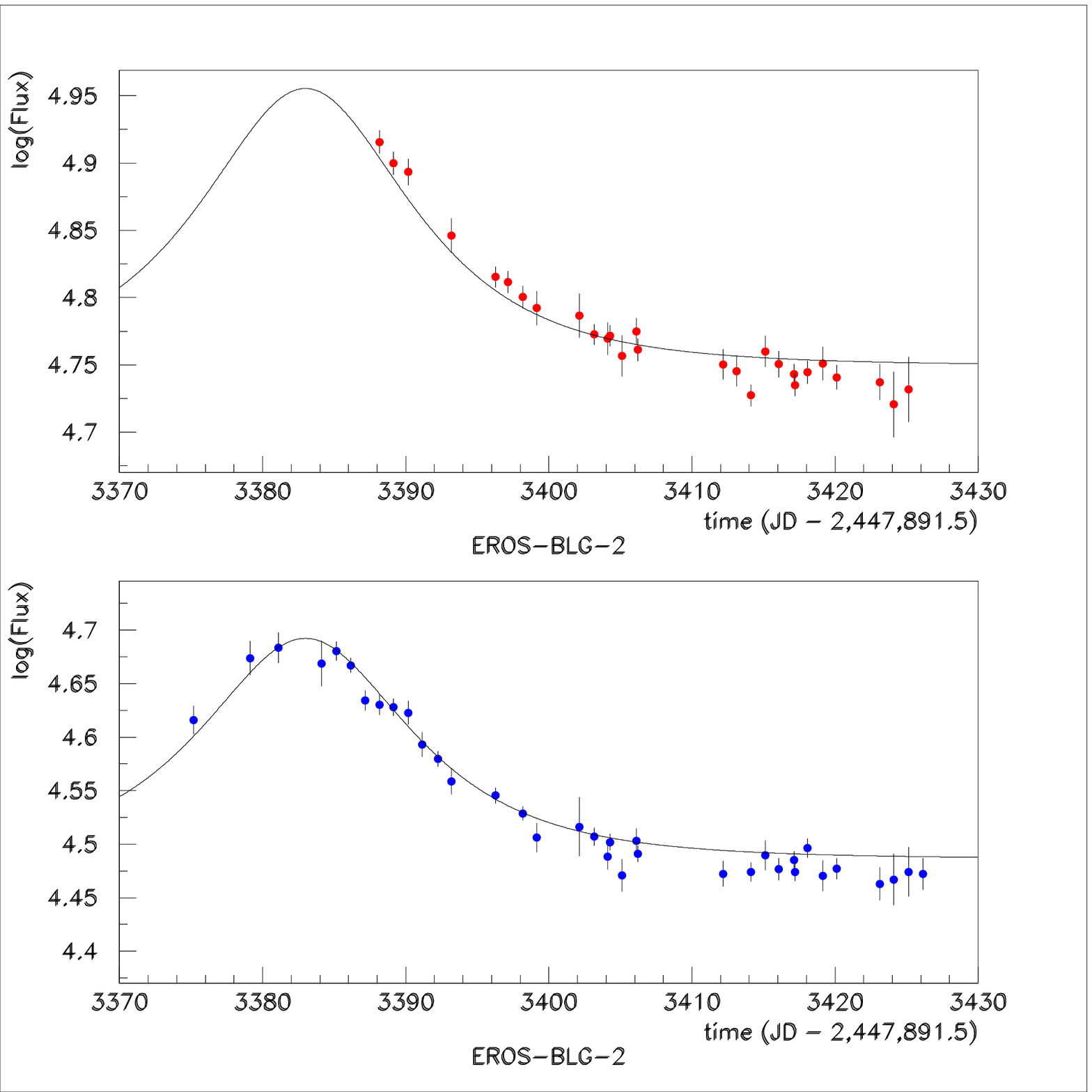}
 \end{minipage}
 \begin{minipage}[c]{.49\textwidth}
 \includegraphics[width=7.5cm]{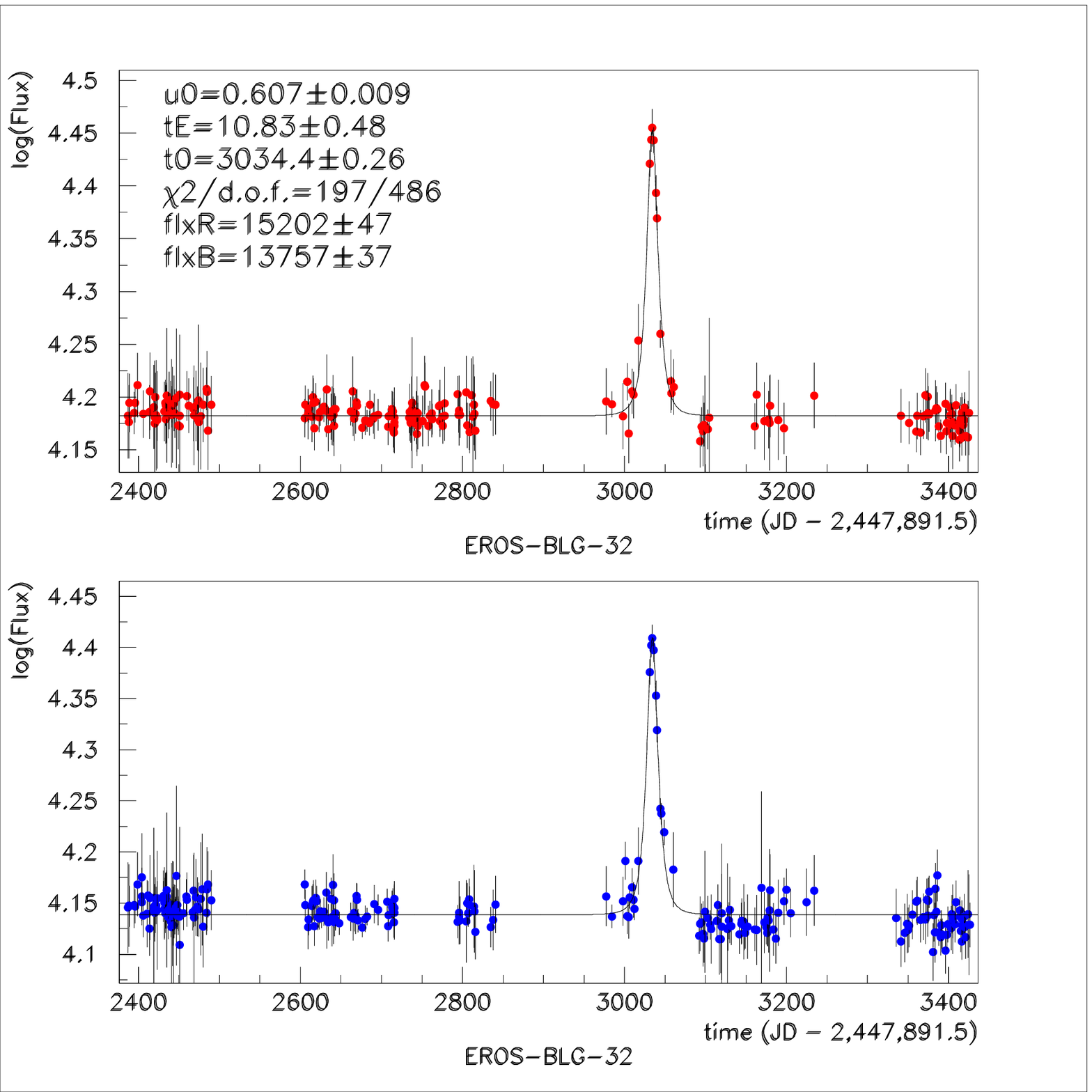}
\end{minipage}
\hspace{0.2cm}
\begin{minipage}[c]{.49\textwidth}
\includegraphics[width=7.5cm]{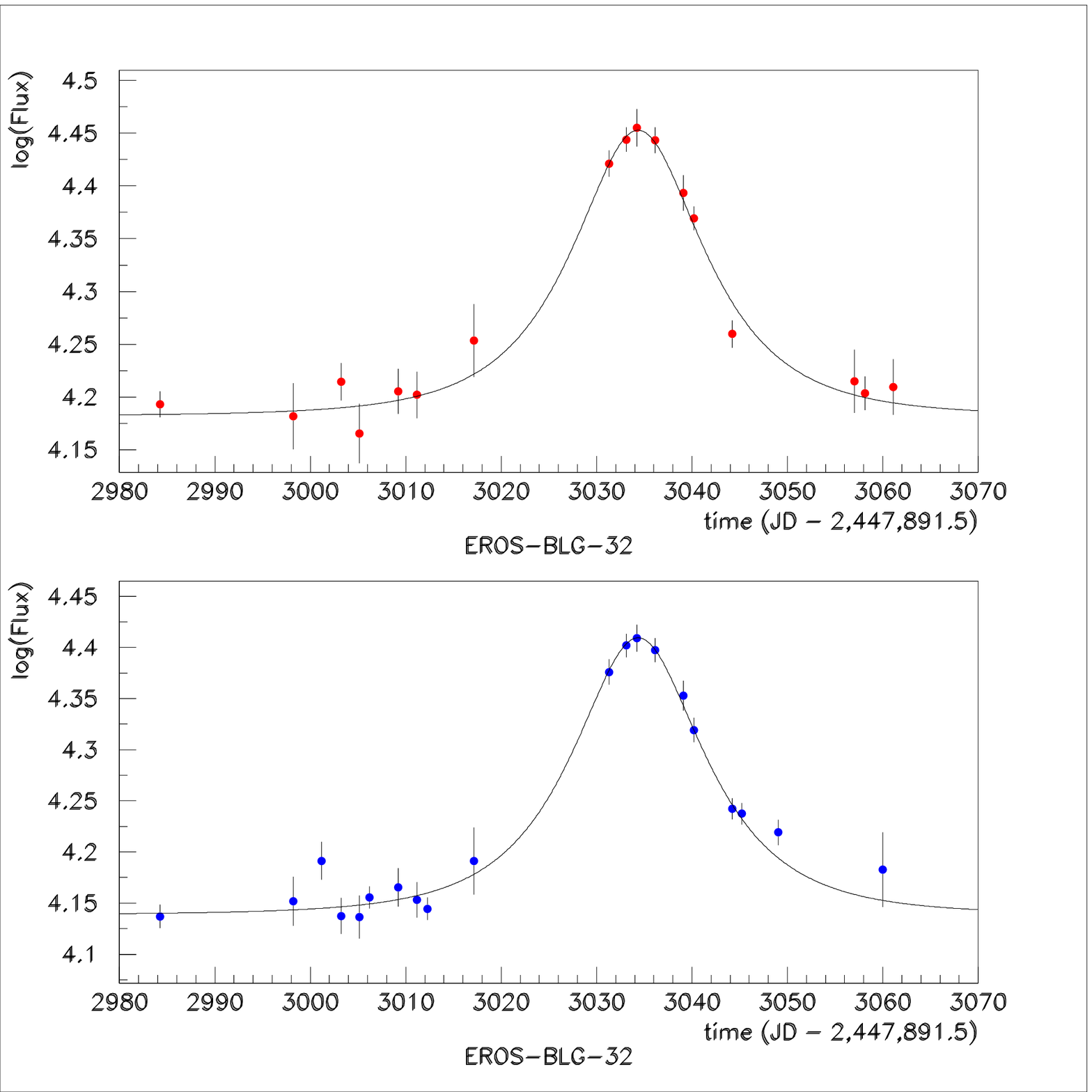}
\end{minipage}
\caption{The light curves of the EROS~2 microlensing candidates
   \#4 to \#6 (see Table~\ref{tab:cand}). In each box the upper
   light curve refers to the EROS red filter and the lower light curve
   to the EROS blue filter. Full span of the light curves is shown in
   the left column and corresponding zoomed light curves are in the
   right column. The 5 parameters obtained by the fit of the
   Paczy\'nski profile are shown (on full span only), as well as the
   $\chi^2$ values of the fit.}  
  \label{fig:cdl_candidates1}
\end{figure}

   Fig.~\ref{fig:cdl_candidates} to \ref{fig:cdl_candidates5} show
   the light curves for the 16  
events. In Table~\ref{tab:cand} we present the characteristics of the 16
microlensing candidates with clump-giant sources (of the extended
  clump area) and $u_0<1$.      
The mean and  standard deviation of the time scales distribution
   (see Fig.~\ref{fig:distri_evt_te}) for these events are
\begin{eqnarray}
\langle t_\e\rangle & = & 33.3 \;\rm days\\
\sigma(t_\e) & = & 39.6 \;\rm days\;. 
\label{eqn:meantime}
\end{eqnarray}

\begin{figure}[h]
 \centering
 \includegraphics[width=7.8cm]{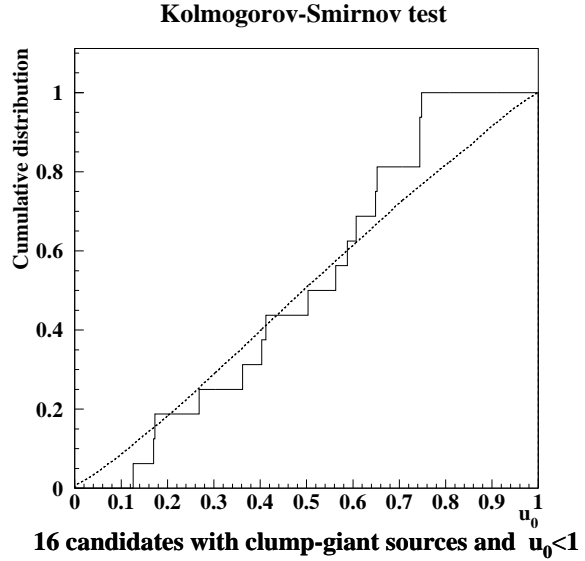}
 \caption{Kolmogorov-Smirnov test for the impact parameter of the 16
   candidates with clump-giant sources (of the extended clump
     area) and $u_0<1$. The maximal 
   distance between the experimental cumulative distribution of $u_0$
   (solid line) and the expected one (dashed line) is $D_{\rm max}$=0.23. 
   This yields a Kolmogorov-Smirnov probability $P_{KS}(D_{\rm max})$=34\%.}
  \label{fig:testks_u0}
\end{figure}

In order to check whether the experimental distribution of the
observed impact parameters are drawn from the same distribution as the
one expected for microlensing events, we use a Kolmogorov-Smirnov
test. The theoretical cumulative distributions are calculated by
selecting the Monte Carlo (MC) simulated events (generated randomly, see
\S\ref{section:efficiencies}) with the same order of time scales as the
observed ones and that were chosen by our analysis cuts. This method
takes implicitly into account the detection efficiency, which will be
presented in the next section. Fig.~\ref{fig:testks_u0} shows the
cumulative distribution of the impact parameters for the 16 candidates
with clump-giant sources (of the extended clump area) and
$u_0<1$. The dotted line 
refers to the expected $u_{\rm 0MC}$ distribution for microlensing. The
Kolmogorov-Smirnov probability $P_{\rm KS}$ indicates the significance of
the similarity of two distributions at distance $D_{\rm max}$ from each
other. We obtain $D_{\rm max}=0.23$ which corresponds to $P_{\rm KS}=34\%$,
which shows a good agreement between the measured and expected
distributions.  

\section{Detection efficiency}
\label{section:efficiencies}

To determine the optical depth (see Eq. \ref{eqn:theo_opt}), we first  
evaluate the detection efficiency for each field as a function of time scale by
using Monte Carlo simulated light curves. We superimpose
artificial microlensing events, with randomly generated parameters (impact
parameter, date of maximum amplification and time scale), on 
each of the real monitored light curves, and find the fraction that are 
recovered by our detection algorithm. Thus, the detection efficiency
is given by 

\begin{equation}
\epsilon(t_{\rm EMC}\;\in\;bin\;i)=\frac{N_{\rm
    DE}(t_\e\;\in\;bin\;i)}{N_{\rm GE,
    u_{0MC}<1}(t_{\rm EMC}\;\in\;bin\;i)} 
\label{eqn:eff}
\end{equation}   
where $t_{\rm EMC}$ is the generated time scale, $N_{\rm DE}$ is
the number of simulated events detected by our analysis, $t_\e$
is the time scale obtained by the microlensing fit, and
$N_{\rm GE,u_{0MC}<1}$ is the number of generated events with an impact
parameter $u_{\rm 0MC}<1$.  

\begin{figure}[h]
 \centering
 \includegraphics[width=7.8cm]{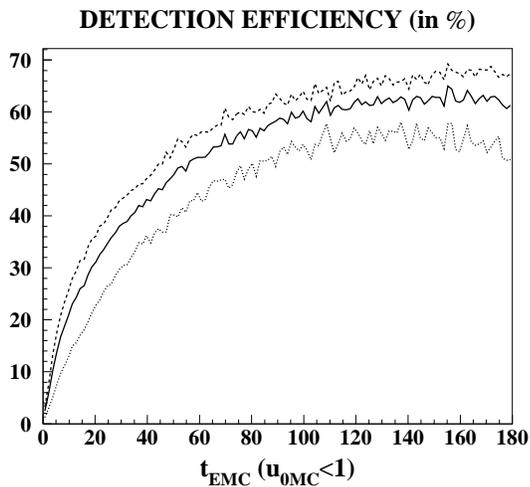}
 \caption{Detection efficiency as a function of the event time scale
   (in days) averaged over all 15 fields (solid line) and 
   two sub-groups of 10 fields (dashed line) and 5 fields (dotted
   line), with different time sampling: $\sim~350$ and
   $\sim~180$ data points respectively.} 
  \label{fig:eff_cg2_moy_1}
\end{figure}

      The microlensing parameters of the simulated events are drawn uniformly: 
the impact parameter $u_{\rm 0MC}$ in the interval [0,2] and the date
of maximum amplification $t_{\rm 0MC}$ in the observation period, with
a margin of 180 days before and after respectively the first and last
day of the observations $[T_{\rm first}-180,T_{\rm
  last}+180]$. The time period for the detection efficiency
  determination, equal to 1418 days, corresponds to the observation
  period (1058 days) extended by a 180 days margin on both sides, in
  order to check whether we are sensitive to microlensing events with
  maximum magnification occurring just before or after the actual observation
  period. Finally the Einstein ring radius crossing time $t_{\rm EMC}$
is drawn uniformly from a $log(t_{\rm EMC})$ distribution (to enhance
the efficiency precision at small time scales) over the interval
[1,180] days. Efficiencies were calculated only until $t_\e=180$ days
because there were no events detected longer than 145
days. Fig.~\ref{fig:eff_cg2_moy_1} shows these efficiencies
  averaged over two sub-groups of 10 fields (solid line)
and 5 fields (dashed line), as well as the global detection efficiency
which is the average over all 15 fields (bold line). These sub-groups refer to the most and
least densely sampled light curves, with $\sim$ 350 data
points and $\sim$ 180 points respectively within the observation
period, which is the same for all fields (1058 days). 
For the high signal to noise events used in this paper, the
efficiency is affected mostly by time gaps in the data. For example,
for long events with $t_\e\sim100$ days, the 60\% efficiency reflects 
the non-observability of the galactic center during the southern
summer. Shorter time scale events are affected by periods of bad
weather and instrumental failures.

\section{Optical Depth}
\label{section:optical_depth}
   
   The microlensing optical depth can be defined as the probability
that a given star, at a given time $t$, is magnified by at least 1.34,
i.e. with an impact parameter $u(t)<1$. The optical depth is then given
by   

\begin{equation}
\tau = {\pi\over 2 N_\star T_{\rm obs}}\sum_{i=1}^{N_{\rm ev}}{t_{\rm E,i}\over
  \epsilon(t_{\rm E,i})}\;, 
\label{eqn:theo_opt}
\end{equation}
where $N_\star$ is the number of monitored stars, $T_{\rm obs}$ is the
observation period, $t_{\rm E,i}$ is the measured Einstein ring radius
crossing time of the $i$th candidate and $\epsilon(t_{\rm E,i})$ is the
corresponding global detection efficiency (see
Fig.\ref{fig:eff_cg2_moy_1}). Note that the above 
expression for $\tau$ only applies to objects whose mass and velocity
cause events in the time scale range with significant
efficiency. There could be more optical depth from events outside this
range.

In Table~\ref{tab:eff_te} we summarize
the time scales of the 16 microlensing candidates with clump-giant 
sources (of the extended clump area) and $u_0<1$, and the
detection efficiencies for each measured 
$t_\e$. 

\begin{table}[!h]
\begin{center} \begin{tabular}{l l c c} \hline\hline
{\em Candidate} & Name & $t_\e$ (days) & $\epsilon(t_\e)$ (in \%) \\ \hline
\#1 & EROS-BLG-16 & 4.7  & 14.2 \\ 
\#2 & EROS-BLG-35 & 8.5  & 21.0 \\ 
\#3 & EROS-BLG-3 & 9.3  &  22.3 \\
\#4 & EROS-BLG-28 & 10.0 & 23.4 \\ 
\#5 & EROS-BLG-2 & 10.4 & 23.8 \\ 
\#6 & EROS-BLG-32 & 10.8 & 24.7 \\ 
\#7 & EROS-BLG-13 & 13.1 & 27.0 \\ 
\#8 & EROS-BLG-33 & 15.7 & 29.1 \\ 
\#9 & EROS-BLG-31 & 18.3 & 32.0 \\ 
\#10 & EROS-BLG-23 & 20.6 & 34.0 \\ 
\#11 & EROS-BLG-11 & 30.3 & 40.3 \\ 
\#12 & EROS-BLG-5 & 35.6 & 43.0 \\ 
\#13 & EROS-BLG-18 & 35.8  & 43.0 \\ 
\#14 & EROS-BLG-4 & 56.2 & 54.7 \\ 
\#15 & EROS-BLG-29 & 108.3 & 63.4 \\
\#16 & EROS-BLG-12 & 145.6 & 62.8 \\

\end{tabular}
\caption{The Einstein ring radius crossing time $t_\e$ and corresponding
  detection efficiency are shown for the 16 microlensing
  candidates with clump-giant sources (of the extended clump area) and $u_0<1$.} 
\label{tab:eff_te}
\end{center}
\end{table}

\begin{figure}[h]
 \centering
 \includegraphics[width=7.8cm]{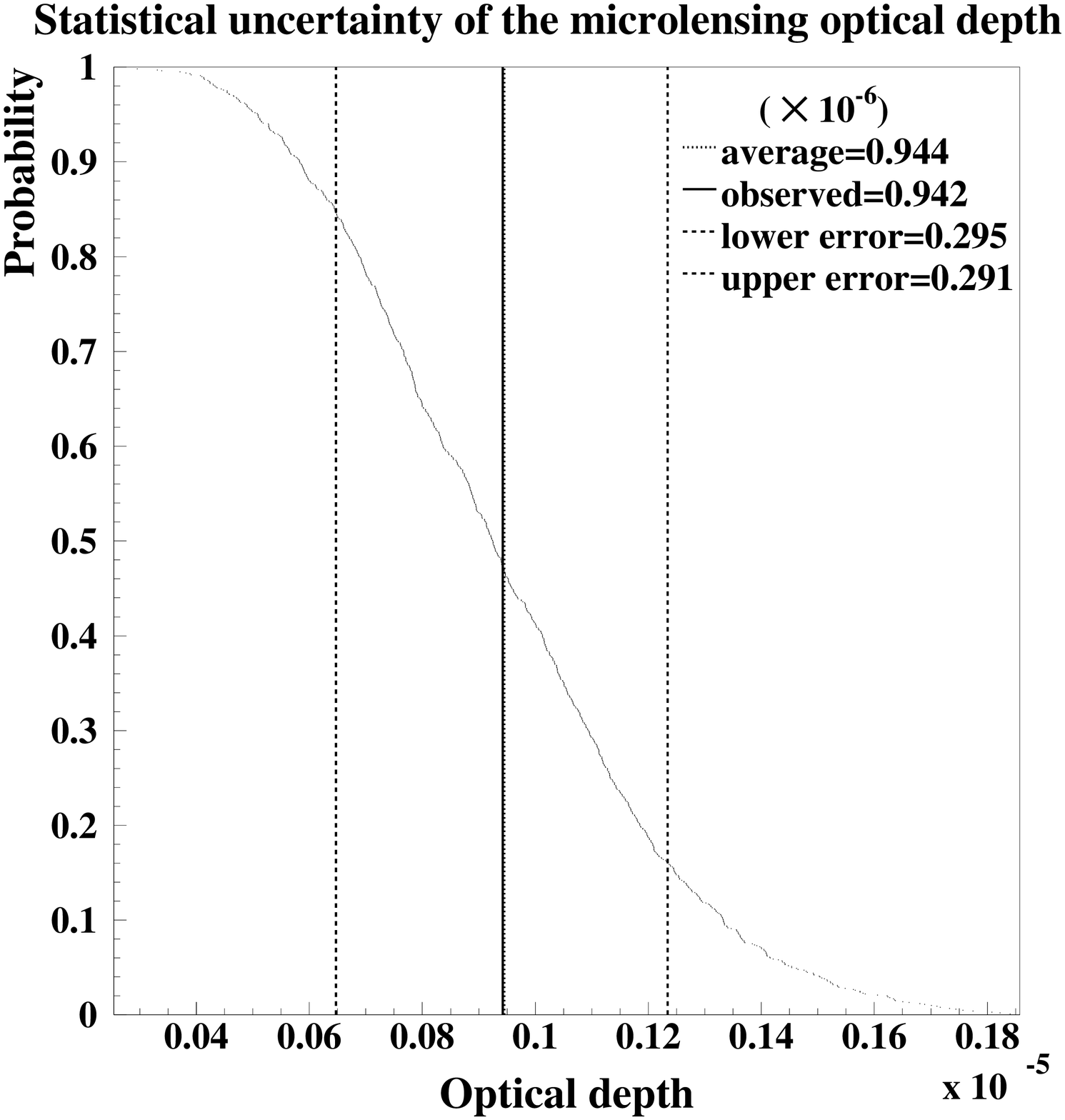}
 \caption{The cumulative distribution of the statistical microlensing optical
   depth drawn from a significant number of virtual experiments for
   our sample of 16 microlensing candidates. The optical depth
   uncertainties are then given by the $\pm1\sigma$ confidence limits
   (dashed lines).} 
  \label{fig:err_profopt_err}
\end{figure}

Fig.~\ref{fig:distri_evt_te} shows the time scale distribution of the raw
counts (dashed line) and corrected for efficiency (solid line), a
rescaling factor having been applied so that the histograms have the same
area. For the derivation of the optical depth we replace the
parameters of Eq.~(\ref{eqn:theo_opt}) by the corresponding values:  
$N_\star=1.42\times 10^6$, equal to the number of clump giants (of the
extended clump area),  
$T_{\rm obs}=1418$ days which corresponds to the actual time period of
the generation of simulated events (for the detection efficiencies
determination, see \S\ref{section:efficiencies}), and finally the
time scales and efficiencies found in Table~\ref{tab:eff_te}. In the
case of the 2 events 
affected by parallax, we considered the time scale obtained when taking
into account this effect, but used the efficiencies for the time scales $t_\e$
determined from a simple microlensing fit without parallax, as
initially found by our analysis. Regarding the events with blending, 
we used the time scale uncorrected for this effect, 
otherwise we would have had to estimate the number of blended unseen stars to
add it to our optical depth equation. We have checked that the
measured optical depth depends very little on these assumptions. 
Moreover, as we will justify below in a study to quantify the effect
of blending on bright stars, these are on average unaffected. We
obtain a bulge microlensing optical depth of
 
\begin{equation}
\tau = 0.94^{+0.29}_{-0.30} \times 10^{-6} \;\;at\;\;
(l,b)=(2.\hskip-2pt^\circ 5,-4.\hskip-2pt^\circ0) \; .
\label{eqn:obs_opt}
\end{equation}
Note that the optical depth, and the associated errors, are valid only for
objects within the range of detection $\sim 2\;{\rm
  days}<t_E<180\;{\rm days}$. 
The $(l,b)$ position is an average of positions of the clump giants
(of the extended clump area) in
the 15 fields. The uncertainties are statistical, estimated using the
same technique as in Alcock~et~al.~(2000). To do so, a significant number of
experiments were simulated. For each experiment we generated
the number $n$ of ``observed'' microlensing events, according
to Poisson statistics with a mean of $\mu=16$, equal to the
number of actually observed candidates. To each of the $n$ events, one
of the 16 measured time scale was assigned randomly (being uniformly
drawn), thus obtaining an optical 
depth estimate for each virtual experiment. The uncertainties are then
given by the $\pm1\sigma$ values from the average of the simulated
optical depth distribution (see Fig. \ref{fig:err_profopt_err}). The
2$\sigma$ can be calculated in the same way, yielding $\tau =
0.94^{+0.68}_{-0.46} \times 10^{-6}$. 

The errors can also be estimated analytically (\cite{HAN95b})

\begin{equation}
\sigma(\tau) = \tau\frac{\sqrt{<t_\e^2/\epsilon^2>}}{<t_\e/\epsilon>}
\frac{1}{\sqrt{N_{\rm ev}}}=
0.29\times 10^{-6}\; ,
\end{equation}
in very good agreement with the uncertainties given in
Eq. ~(\ref{eqn:obs_opt}).

\begin{figure}[h]
 \centering
 \includegraphics[width=7.8cm]{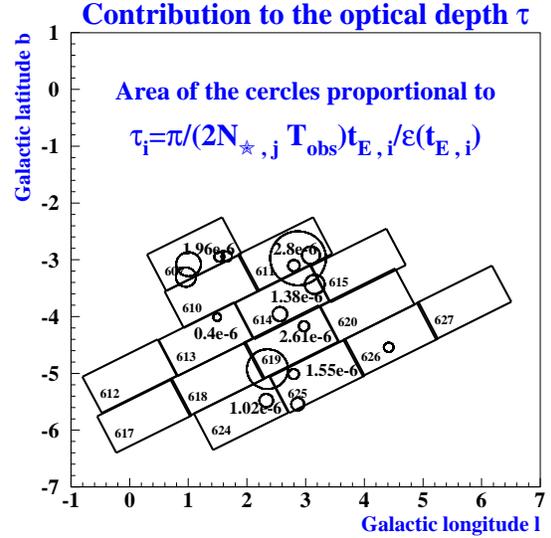}
 \caption{The contribution of the 16 candidates to the observed
   optical depth. The area of the circles is proportional to the
   optical depth due to each microlensing candidate. The measured optical depth
   in each field is also shown, as well as the number of
   the EROS fields.} 
  \label{fig:cont_candidates}
\end{figure}

The contribution of each of the 16 candidates to the measured optical
depth is shown in Fig.~\ref{fig:cont_candidates}, where the area
of the cercles is proportional to the individual optical depth due to
each event 
$\tau_i=\pi/(2N_{\star,\rm j}T_{\rm obs})t_{\rm E,i}/\epsilon(t_{\rm E,i})$, 
$N_{\star,\rm j}$ being the number of stars in field $j$, with the shortest
event lasting $\sim5$ days and the longest $\sim146$ days. We also
show the measured optical depth in each field.  

\section{Effect of blending on the measured optical depth}
\label{section:effect_blending}

In order to check that the measured optical depth given
by microlensing events with clump-giant sources is not
significantly affected by blending, we created a set of artificial images 
with simulated microlensing events, and calculated the optical
depth from the candidates detected by our selection
pipeline on the simulated light curves. Two types of synthetic
images, corresponding to the two EROS passbands (see
\S\ref{section:data}) and with a size of $512\times 512$ pixels,     
were generated from a catalog derived from the Holtzman et al. (1998)
luminosity function in the Baade Window. The catalog contained 365,000
stars which were placed randomly over the $512\times 512$ pixels
area. The faintest catalog star was 8 magnitudes dimmer than the faintest
reconstructed star considered in this paper. 
 
   On an arbitrarily selected synthetic reference image, 20\% (73,000) of
the total number of artificial stars were chosen to be microlensed. We
then generated in each color a sequence of 3,600 images equally
spaced in time, the unit of time being 1 image. On each ensemble
of $2\times20$ images (blue and red filters), about 400 microlensing
events were generated. To avoid photometric
interference between simulated events, only stars at least 20 pixels
away from each other and with similar magnitudes were lensed. The
microlensing events were generated with impact parameter   
$u_0$ randomly drawn between 0 and 1.5, date of maximum
amplification $t_0$ equal to the center of the ensemble with a margin
of 0.5 images, and time scale $t_\e=$ 5 images. For the microlensing
fit we used the ensemble containing the fluctuation, plus an
additional 20 images generated with no events in order to determine
the baseline.      

\begin{figure}[h]
 \centering
 \includegraphics[width=7.8cm]{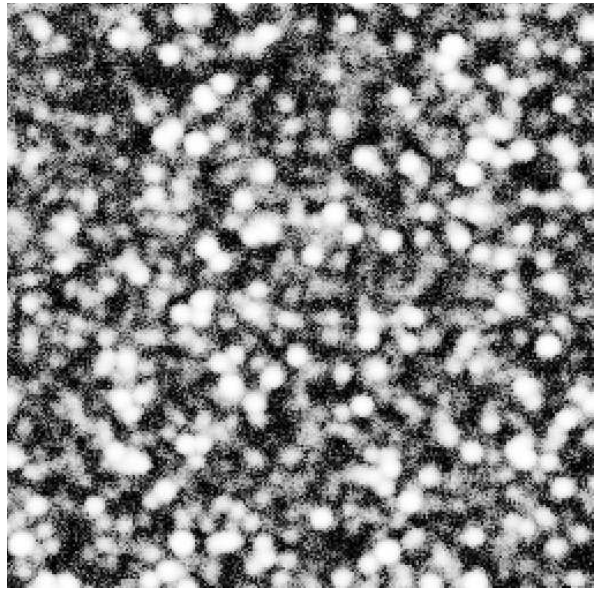}
 \caption{Example of a synthetic image as described in
   \S\ref{section:effect_blending} (256$\times$256 pixels are shown).}    
  \label{fig:mcbref}
\end{figure}

\begin{figure}[h]
 \centering
 \includegraphics[width=7.8cm]{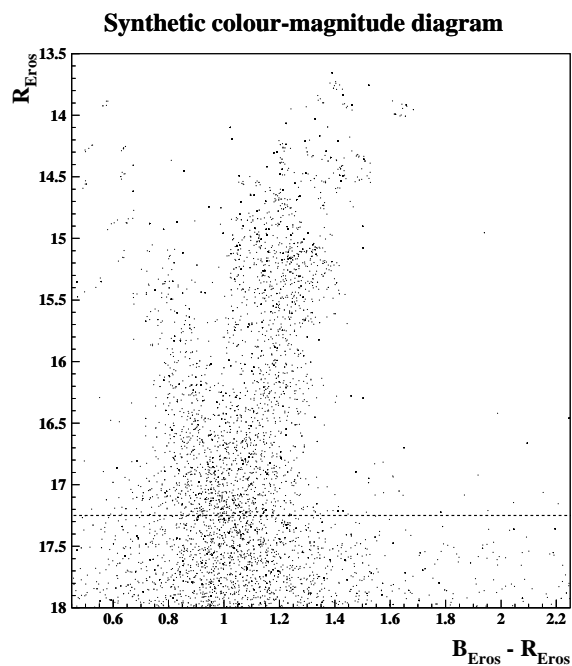}
 \caption{Artificial colour-magnitude diagram. The dashed line
   indicates the magnitude cut for the selection of bright stars.}
  \label{fig:fake_cmd}
\end{figure}

   Roughly 10,000 stars of the total number of artificial stars were
reconstructed by our software on each synthetic image, an example of which
is shown in Fig.~\ref{fig:mcbref}. To define a sample of bright stars
a magnitude cut was performed on the CMD of the synthetic reference
image (see Fig.~\ref{fig:fake_cmd}). A total of 2270 stars were selected, 
corresponding closely to the mean density of bright stars
reconstructed on real EROS images. The analysis pipeline was then
applied to the simulated light curves of this sample of reconstructed
bright stars. 

   A total of 411 generated microlensing events were found
with an impact parameter $u_0<1$ and an average of reconstructed
parameters $<t_\e> = 3.55$ images and $<u_0> = 0.56$. From these
events, 255 were due to the main star, i.e. the brightest
catalog input star in the two pixels around the reconstructed star,
with recovered $<t_\e> = 4.67$ images and $<u_0> = 0.49$. The remaining 156 
events are due to the fainter, blended, component of the reconstructed
star. The average of the recovered time scales for these blended events
is $<t_\e> = 1.72$ images, clearly underestimated, and $<u_0> = 0.66$,
overestimated.

   In the absence of blending and with perfect photometric
resolution, the number of simulated microlensing events one would
expect to recover is $2270\times0.2/1.5=302$, where 2270 is the number of
reconstructed bright stars selected by the magnitude cut in the CMD and
0.2 is the fraction of catalog stars microlensed with impact
parameters less than 1.5. The optical depth being proportional to the
product of the number of events passing the microlensing selection
criteria and their mean $t_\e$, the ratio $R$ of the recovered optical
depth with the generated one yields   

\begin{equation}
R = \frac{411}{302}\frac{3.55}{5.0} = 0.97
\end{equation} 
where $411$ is the number of simulated microlensing events found by
our analysis pipeline, the value 3.55 is the average of the recovered
time scales, 302 is the number of simulated microlensing events one would
expect to recover and 5.0 the average of the input time scales. The error
of the ratio $R$ is estimated to be 10\%. This figure is based on the
statistical error and from small differences in results obtained by
varying within reason the form of the PSF used to generate synthetic
images. The recovery of 97\% of the generated optical depth is a
reassuring result. Thus, our conclusion is that we can neglect
blending effects on the optical depth inferred from microlensing
events with bright source stars.  

\section{Searching the alerts and microlensing events of the MACHO and
  OGLE collaborations in the EROS data}
\label{section:check_macho_ogle} 

In view of our low measured optical depth compared to other
determinations (see Table~\ref{tab:opt}), it is important to check
that microlensing events had not been lost in the analysis procedure
in unsuspected ways that are not taken into account by the Monte Carlo
detection efficiency calculation.  To do this, we looked for Galactic
Center events that had been found independently by the MACHO and OGLE
collaborations within our observation period in the 15 fields we
analyzed, and whose magnitudes are brighter than our cut in the CMD (see
\S\ref{section:candidates}). We also looked for alerts found
by the EROS trigger.

    From the MACHO collaboration, we considered the 211 online
alerts\footnote{http://darkstar.astro.washington.edu} and 99 
published events (\cite{ALC00}) found by differential photometry.
From the OGLE collaboration, we considered the 89 alerts reported 
during the years 1998 and
1999\footnote{http://www.astrouw.edu.pl/$\sim$ogle/ogle2/ews/ews.html} and
the 214 candidates published by Udalski et al. (2000).  Regarding the
EROS~2 alert
system\footnote{http://www-dapnia.cea.fr/Spp/Experiences/EROS/alertes.html},
although it was only operational after May 1999, beyond the
time period of the data analyzed in this paper, a test version was
performed during a limited time yielding three alerts to be considered for the
search. From these five sources, a total of 22 events occurred within
the observation period considered in this paper (from July 1996 to 31
May 1999) and concerned stars bright enough to pass our magnitude cut.

      Of these 22 events, 13 were identified by our analysis
pipeline as microlensing candidates, 8 of which have clump-giant
sources and an amplification $A_0>1.34$. The 9 remaining events were
not found. Two were rejected by our selection criteria: one
because of excessive fluctuations outside the amplification peak and
the other event because the improvement of a microlensing fit over a
constant-flux fit was not good enough. Another two events occurred on
source-stars that do not appear in the EROS catalog and, as such,
cannot be considered for the optical depth measurement. These two stars
have a magnitude at the limit of our magnitude cut and are at the edge
of brighter stars, which explains their non-appearance in the
catalog. Finally, 5 events occured during periods that were at best 
sparsely sampled by EROS due to bad weather or technical
problems. Their non-detection is thus normal and corrected for by our
Monte Carlo detection efficiency computation. 

    Note that the optical depth estimate presented
in this paper is unaffected by these results, since none of the
``unseen'' MACHO and OGLE candidates were not found without a
supporting reason (i.e., the analysis pipeline behaved like we expected it
to).

\section{Discussion and conclusion}

\begin{table*}
\begin{center} \begin{tabular}{l l c c c c c} \hline \hline
   & Observed & $l,b$ & {\scriptsize Optical depth}  & No. & No. & Bulge\\
  Group & optical depth & & {\scriptsize Baade Window} & of & stars & seasons \\
  & ($\times 10^{-6}$) & $(^\circ)$ & ($\times 10^{-6}$) & events &
  ($\times 10^6$) &\\
  &             &       & $\tau\pm1\sigma$  & & &\\ \hline

1.~EROS~2 & $\tau_{bulge}=0.94^{+0.29}_{-0.30}$ & 
  $2.5,-4.0$ & $1.08\pm0.30$ & 16 CG & 1.42 & $\sim$ 3 \\ 
2.~MACHO & $\tau_{bulge}=3.90^{+1.8}_{-1.2}$ & 
  $2.6,-3.6$ & $3.86\pm1.50$  & 13 CG & 1.3 & 190 days  \\ 

\multirow{2}{12mm}{3.~MACHO} & $\tau=2.43^{+0.39}_{-0.38}$ &
\multirow{2}{12mm}{2.7,-3.4}  &  & \multirow{2}{5mm}{99} &
\multirow{2}{5mm}{17} & \multirow{2}{7mm}{$\sim 3$}\\  
& $\tau_{bulge}=3.23^{+0.52}_{-0.50}$ & & $3.11\pm0.51$ & & & \\ 

4.~MACHO & $\tau_{bulge}=2.0\pm0.4$ & $3.9,-3.8$  &
$2.13\pm0.40$ & 52 CG & 2.1 & $\sim$ 5\\ 
5.~OGLE & $\tau=3.30\pm1.2$ & $1,-4$ &
$3.3\pm1.20$  & 12 & $\sim 1$ & $\sim$ 3\\ 
6.~OGLE~II & - & - & - & 214 & 20.5 & $\sim$ 3 \\
\hline
\end{tabular}
\caption{Microlensing optical depth estimations at the Baade Window
  ($l=1^\circ,b=-4^\circ$), by the EROS~2, MACHO and OGLE collaborations : 
1.~this paper, 2.~\cite{ALC97}, 3.~\cite{ALC00}, 4.~\cite{POP00},
5.~\cite{UDA94b}, 6.~\cite{UDA00}.}
\label{tab:opt}
\end{center}
\end{table*}

        The optical depth obtained above (Eq. \ref{eqn:obs_opt}) is
low compared to other determinations, as can be seen in
Table~\ref{tab:opt}. For direct comparison among these experiments,
we also report in this Table the observed optical depths extrapolated to the
Baade Window position ($l=1^\circ,b=-4^\circ$), after applying an
optical depth gradient in the $l$ and $b$ directions. We deduced a 
rough estimate for the gradient: $\partial\tau/\partial b = 0.45\times
10^{-6} deg^{-1}$ and $\partial\tau/\partial l = 0.06\times
10^{-6} deg^{-1}$, from several microlensing maps
predicted by various non-axisymmetric models
(\cite{HAN95b}, \cite{ZHAO96}, \cite{BISS97},
\cite{EVA02}). The expected opticals depths, for these models, over the interval of
  Galactic longitude and latitude of our fields ($-6>b>-2, 6>l>0$) 
 ranges roughly from $\tau \sim 1.8 \times 10^{-6}$ to $\tau \sim 0.6
 \times 10^{-6}$, as one goes farther away from the 
  Galactic Center. For comparaison with the range of the measured
  opticals depths in the EROS fields see Fig.\ref{fig:cont_candidates}.

The first conclusion that can be drawn is that the quoted measurements
are consistent with our optical depth estimate only at the 
$2\sigma$ level. Moreover, the predicted optical depths seem to be
more in agreement with our value. Indeed,  
$\tau_{bulge}\sim1.3\times 10^{-6}$ is expected at the Baade Window by
Han \& Gould (1995b), $\tau_{bulge}\sim0.8-0.9\times 10^{-6}$ is the inferred
estimation by Bissantz et al. (1997) at the same position, and
the predicted optical depths by Evans \& Belokurov (2002) with two
different models are $\tau_{bulge}\sim1\times 10^{-6}$ and
$\tau_{bulge}\sim1.5\times 10^{-6}$, although a third model of these
authors gives a higher estimate
$\tau_{bulge}\sim2\times 10^{-6}$. All of the above mentioned models consider
a barred non-axisymmetric bulge. The MACHO and OGLE optical
depth measurements are systematically higher than the predicted values,
except for the Popowski et al. (2000) determination which is more in
agreement with the models. Furthermore, recently Binney, Bissantz \&
Gerhard (2000) argued that an optical depth for bulge sources as large
as the ones inferred by the MACHO collaboration (\cite{ALC97},
\cite{ALC00}) is inconsistent with the rotation curve and the local
mass-density measurements.  

        We report 3 microlensing candidates with long durations,
$t_\e>50$ days: $t_\e=56, 108, 146$ days, all in different fields. These
events contribute about 30\% to the optical depth. Long time scale 
events, difficult to reconcile with the known mass functions, were
already present in the bulge clump-giant sample from Alcock et
al. (1997). They found 3 candidates with $t_\e>75$ days. It was
suggested that they 
might be due to stellar remnants (\cite{ALC97},
\cite{HAN95a}) or to directions where there is a spiral arm
concentration (\cite{DER99}, \cite{PEA99}). Popowski et al. (2000) also
reported 10 long events, with $t_\e>50$ days, contributing 40\% to
the measured optical depth, half of them being concentrated in one
field. In addition, the Einstein ring radius crossing-time 
distribution of the 214 microlensing candidates found by the OGLE
collaboration (\cite{UDA00}), has the same type of tail toward
long time scales ($t_\e>50$ days) as the distributions found by MACHO,
although they are not concentrated in particular fields but rather 
uniformly scattered. Recently, Evans \& Belokurov (2002)
pointed out that bar streaming increases significantly the
amplification durations, with a growing gradient in the mean
time scales from the near-side to the far-side of the bar.    

        In our view, the most robust way to resolve the optical depth
issue, reconciling Galactic structure with microlensing
observations, is to obtain a larger sample of clump-giant events.  We
expect to increase our sample of candidates by a factor 5 by the time
EROS shuts down in 2002. From the preliminary work of Popowski et
al. (2000), one may expect the MACHO sample to be increased by a factor
1.3. Moreover, the OGLE data   
set represents a potentially rich source of additional
events. In the future, the coming of new survey telescopes such as VST and
VISTA, will enhance the possibility to distinguish between Galactic
models, especially if microlensing observations are done in the $K$ band
in the inner $5^\circ\times5^\circ$ region of the Galactic Center
(\cite{GOU95a}, \cite{EVA02}). Thus, the prospects for clarifying this
question over the next few years are very promising.

\bigskip

\begin{acknowledgements}
  We are grateful to D.~Lacroix and the technical staff at the Observatoire
  de Haute Provence and A.~Baranne for their help in refurbishing
  the MARLY telescope and remounting it in La Silla. We are also
  grateful to the technical staff of ESO, La Silla for the support
  given to the EROS project. We thank J-F.~Lecointe and A.~Gomes for the
  assistance with the online computing. Work by A.~Gould was supported
  by NSF grant AST~02-01266 and by a grant from Le Centre Fran\c{c}ais pour
  L'Accueil et Les \'Echanges Internationaux. Work by C.~Afonso was
  supported by PRAXIS XXI fellowship-FCT/Portugal.  
\end{acknowledgements}

\clearpage
\onecolumn

\begin{table}[!h]
\begin{center}
\setlength{\tabcolsep}{0.7mm} 
\begin{tabular}{l l c c c c c c c c} \hline\hline
\multicolumn{2}{c}{EROS~2 candidates} & $\alpha$ (J2000) & $\delta$
(J2000) & $R_{\rm EROS}$ & $B_{\rm EROS}$ & $t_{0}$ &
$t_{\e}$ & $A_{0}$ & $\chi^2$/d.o.f. \\ \hline \hline
\#1 & EROS-BLG-16 & 18:04:57.5 & -29:40:9.8 & 16.1 & 17.3 & 2714.62$\pm$0.12 & 4.64$\pm$0.24 & 1.77 & 199/551\\ \hline

\#2 & EROS-BLG-35 & 18:13:26.7 & -27:21:53.9 & 16.2 & 17.3 & 3017.83$\pm$0.29 & 8.46$\pm$0.31 & 1.91 & 216/261 \\ \hline

\#3 & EROS-BLG-3 & 18:00:46.4 & -29:06:55.7 & 16.3 & 17.9 & 3004.18$\pm$0.28 & 9.30$\pm$0.47 & 1.77 & 295/541 \\ \hline

\#4 & EROS-BLG-28 & 18:08:51.3 & -28:27:11 & 16.2 & 17.6 & 2631.01$\pm$0.06 & 10.04$\pm$0.12 & 2.63 & 862/460\\
 & {\scriptsize MACHO 97-BLG-05} & & & & & & & & \\ \hline

\#5 & EROS-BLG-2 & 18:01:02.5 & -29:00:11.6 & 15.1 & 16.6 & 3382.99$\pm$0.33 & 10.38$\pm$0.35 & 1.61 & 1169/518 \\ 
& {\scriptsize OGLE 1999-BUL-06} & & & & & & & & \\ \hline

\#6 & EROS-BLG-32 & 18:11:51.5 & -29:00:33.5 & 16.5 & 17.5 & 3034.41$\pm$0.26 & 10.84$\pm$0.48 & 1.87 & 197/486\\ \hline

\#7 & EROS-BLG-13 & 18:04:33.6 & -28:07:32.2 & 15.1 & 16.4 & 3105.74$\pm$0.20 & 13.05$\pm$0.35 & 1.60 & 313/465\\
& {\scriptsize EROS-98-BLG-4} & & & & & & & & \\
& {\scriptsize OGLE 1998-BUL-23} & & & & & & & & \\ \hline

\#8 & EROS-BLG-33 & 18:14:32.5 & -29:14:46.4 & 15.7 & 16.7 & 2646.50$\pm$0.26 & 15.69$\pm$0.26 & 1.60 & 911/448\\ \hline

\#9 & EROS-BLG-31 & 18:12:43.4 & -29:38:28.5 & 16.4 & 17.5 & 2812.56$\pm$0.03 & 18.29$\pm$0.18 & 5.84 & 1681/208\\ 
&  {\scriptsize with blending} &  $C_{\rm R}$ & $C_{\rm B}$ & & &  & & &\\
&  &  0.62$\pm$0.05 & 1.00$\pm$0.08 & 16.9 & 17.6 & 2812.59$\pm$0.04 & 22.31$\pm$1.78 & 8.04 & 406/198\\ \hline

\#10 & EROS-BLG-23 & 18:07:06.4 & -28:42:32.8 & 16.3 & 17.5 & 3071.07$\pm$0.39 & 20.63$\pm$0.39 & 2.58 & 255/522\\
& {\scriptsize EROS-98-BLG-2} &  & & & & & & & \\ \hline

\#11 & EROS-BLG-11 & 18:04:9.7 & -27:44:35 & 15.8 & 17.2 & 2790.94$\pm$0.19 & 30.25$\pm$0.38 & 1.98 & 883/552\\
& {\scriptsize OGLE BUL$\_$SC35-144974} & & & & & & & & \\
& {\scriptsize MACHO 97-BLG-58} &  & & & & & & & \\ 
&  {\scriptsize with blending} &  $C_{\rm R}$ & $C_{\rm B}$ & & &  & & & \\
&  &  0.73$\pm$0.14 & 0.55$\pm$0.11 & 16.2 & 17.8 & 2791.00$\pm$0.20 & 35.46$\pm$3.25 & 2.49 & 715/550\\ \hline

\#12 & EROS-BLG-5 & 18:01:10.2 & -29:48:55 & 16.1 & 17.5 & 3423.02$\pm$0.18 & 35.57$\pm$0.35 & 2.17 & 576/749\\
& {\scriptsize OGLE 1999-BUL-07} &  & & & & & & & \\
& {\scriptsize MACHO 99-BLG-12} &  & & & & & & & \\ \hline

\#13 & EROS-BLG-18 & 18:06:20.4 & -27:56:13 & 16.3 & 17.4 & 2987.59$\pm$0.41 & 35.83$\pm$0.86 & 2.90 & 554/454\\ \hline

\#14 & EROS-BLG-4 & 18:00:6.9 & -29:38:06 & 16.7 & 18.5 & 2743.93$\pm$0.08 & 62.89$\pm$0.32 & 7.88 & 422/774 \\
& {\scriptsize MACHO 97-BLG-26} & & & & & & & & \\ \hline

\#15 & EROS-BLG-29 & 18:10:56.2 & -29:24:24.4 & 14.3 & 15.5 & 3363.14$\pm$0.27 & 116.31$\pm$0.79 & 3.82 & 938/551\\ 
& {\scriptsize with parallax} &   $\phi$ & $\pi_{\rm E}$ & & & & \\
& &  -1.492$\pm$0.567 & 0.229$\pm$0.023 & 14.3 & 15.5 & 3362.90$\pm$0.75 & 108.34$\pm$0.68 & 24.40 & 537/549\\ \hline

\#16 & EROS-BLG-12 & 18:03:53.2 & -27:57:36 & 15.8 & 17.2 & 2491.63$\pm$0.37 & 109.50$\pm$0.75 & 5.93 & 1484/452\\
& {\scriptsize MACHO 96-BLG-12} & & & & & & & & \\ 
& {\scriptsize with parallax} &   $\phi$ & $\pi_{\rm E}$ & & & & \\
& & -0.138$\pm$0.22 & 0.272$\pm$0.063 & 15.8 & 17.2 & 2472.50$\pm$0.94 & 145.61$\pm$2.72 & 8.81 & 552/438\\\hline
\end{tabular}
\caption{The list of the 16 EROS~2 microlensing candidates (with
  clump-giant sources and $u_0<1$). The first column indicates the
  number and name of the EROS~2 candidate. The 
  corresponding MACHO and OGLE candidates/alerts are shown below the EROS
  event name, as well as the EROS alerts, when these have been
  reported. The following two columns (2 and 3) refer to the 
  sky coordinates of the source star. The next columns (4 and 5) show
  the $R_{\rm EROS}$ and $B_{\rm EROS}$ magnitudes of the source
  stars. Columns number 6 and 7 refer to the date of maximum
  amplification $t_0$ and the time scale $t_\e$ of 
  the candidate. The last two columns (8 and 9) indicate the maximum
  amplification $A_0$ of the light curve and the reduced $\chi^2$ of
  the microlensing fit. For candidates \#9 and \#11, the results of the
  microlensing fit taking blending into account are shown. The
  parameters $C_{\rm R}$ and $C_{\rm B}$ refer to the blending
  coefficients for the red and blue light curve (see
  \S\ref{section:event_selection} for blending definiton). Candidates
  \#15 and \#16 are parallax events, the parameters $\pi_\e$ and
   $\phi$ being respectively the amplitude of the displacement in the
   Einstein ring due to the Earth's orbital motion and the phase of
   the displacement.} 
\label{tab:cand}
\end{center}
\end{table}

\clearpage

\clearpage

\clearpage

\clearpage

\clearpage

\clearpage

\end{document}